\title{Combinatorial constructions of intrinsic geometries }

\author{Stanislaw Ambroszkiewicz\footnote{Institute of Computer Science, Polish Academy of Sciences, al. Jana Kazimierza 5, 01-248 Warsaw, Poland, Email: sambrosz@gmail.com}
}

\documentclass[a4paper,12pt]{article}
\usepackage{amssymb}
\usepackage{multicol}
\usepackage{color}
\usepackage{xspace}
\usepackage{enumerate}
\usepackage{hyperref}
\usepackage{graphicx}

\begin{document}
\maketitle 
\pagestyle{headings}

\begin{abstract} 
A generic method for combinatorial constructions of intrinsic  geometrical spaces is presented. It is based on the well known inverse sequences of finite graphs that determine (in the limit) topological spaces. If a pattern of the construction is sufficiently regular and uniform, then the notions of metric, geodesic and curvature can be defined in the space as the limits of their finite versions in the graphs. This gives rise to consider the graphs with metrics as finite approximations of the geometry of the space. 
On the basis of simple and generic examples, several nonstandard and novel notions are proposed for the Foundations of Geometry. They may be considered as a subject of a critical discussion. 
\end{abstract}
\tableofcontents
\sloppy

\section{Intrinsic versus extrinsic geometries }

Geometry means earth measurement.  
Origins of synthetic geometry (axiomatic pure geometry called Euclidean geometry) goes back to ancient Greeks and Euclid. 

\url{https://www.britannica.com/science/Euclidean-geometry} : {\em Euclidean geometry, the study of plane and solid figures on the basis of axioms and theorems employed by the Greek mathematician Euclid (c. 300 BCE).} 

Its fundamental concepts include: point, line, plane, curve, distance, angle, surface, area, volume, manifolds and curvature. 

Cartesian (analytic) geometry emerged in 17th century. 
Using the Cartesian coordinate system, geometric shapes (such as curves and surfaces) can be described by Cartesian equations, i.e. algebraic equations involving the coordinates of the points lying on the shapes. The basis is the set of real numbers  $\mathbb{R}$. The Cartesian plane is defined as  $\mathbb{R}^2$, i.e. the Cartesian product $\mathbb{R}\times \mathbb{R}$. Three dimensional Cartesian space is $\mathbb{R}^3$, and, in general,  $\mathbb{R}^n$ is $n$-dimensional Cartesian space. 

$\mathbb{R}^3$ is considered as ``the right'' representation of ``the space'' we live in, and a model of the Euclidean axioms. Cartesian space $\mathbb{R}^n$ is also called $n$-dimensional Euclidean space. 

Synthetic geometrical approaches to  ``non-Euclidean geometry'' were first presented by János Bolyai and Nikolai Ivanovich Lobachevsky in 18th century where ``hyperbolic'' geometry was described in axiomatic ways. This gave rise to consider different geometries where, contrary to the Euclidean geometry, the parallel postulate was not satisfied. 

Until the middle of the 19th century, geometry was studied from the extrinsic point of view: curves and surfaces were considered as lying in three dimensional Euclidean space $\mathbb{R}^3$ as the ambient space. Starting with the work of Bernhard Riemann, the intrinsic point of view was developed where geometric objects are stand-alone without an ambient space. 
This seems to be more flexible. The ambient space (universe) is not needed. 
For example, from the intrinsic point of view, a cylinder is locally "the same" as the Euclidean plane. 
%
%
Also flat torus and flat Klein bottle can be easily constructed as intrinsic geometries. 
 
Although Riemannian manifolds present intrinsic point of view, their definition is based on the notions of Cartesian spaces and  inner product on the tangent space at each point that varies smoothly from point to point. They are complex mathematical abstractions. Even intrinsic elliptic geometry of 2-sphere, as a Riemannian manifold, has a complex definition.

A hyperbolic $n$-space, denoted $\mathbb{H}_\kappa^n$, where $\kappa < 0$, is the maximally symmetric, simply connected, $n$-dimensional Riemannian manifold with the constant negative sectional curvature $\kappa$. It is a definition. Model of such space is defined in the ambient space $\mathbb{R}^{n+1}$. 

Others known models of hyperbolic space include the Klein model, the hyperboloid model, the Poincaré ball model, and the Poincaré half space model. 
All of them are defined within ambient Euclidean space. 

Most of the other  known hyperbolic manifolds are quotients  of $H_\kappa^n$ by a discrete group 
of isometries on $H_\kappa^n$.

There are simple geometries that are not Riemannian manifolds, like the Taxicab plane. It is the Cartesian product  $\mathbb{R}\times \mathbb{R}$ equipped with the Taxicab metric. Given two points $(x_1,y_1)$ and $(x_2,y_2)$, the distance between them is calculated as  $|x_1 - x_2| + |y_1 - y_2|$. 


It is important to note that $\mathbb{R}$ - the set of real numbers - is the basis for defining Cartesian spaces and   Riemannian manifolds. Actually, $\mathbb{R}$ itself is an abstraction involving an actual infinity. 

The principal goal of our paper is to build the (finitist) foundations of geometry without $\mathbb{R}$; the foundations that are based on inductive constructions of finite structures (graphs with metrics). The intrinsic geometries result as abstractions (limits) of such regular and uniform inductive constructions.

\subsection{Finite approximations } 

It is well known that any compact metric space can be approximated by a sequence of finite polyhedra (i.e. realizations of finite simplicial complexes) since the works of Alexandroff (1937) \cite{Alexandroff} and Freudenthal (1937) \cite{Freudenthal}. 

For a compact metric space $X$ and its open covering $U$, there is a natural notions of graph 
denoted $G(U)$, and nerve denoted $N(U)$. In both cases, the set of vertices is $U$. An edge of the graph $G(U)$ is defined as a pair of elements of $U$ with nonempty intersection. 
Analogously, a simplex of $N(U)$ is represented as a subset of elements of $U$ such that their intersection is 
nonempty. For compact spaces, there are finite refinements of the covering, so that, the corresponding graphs and simplicial complexes are finite structures. 

More and more refined uniform coverings along with some consistency conditions determine an inverse sequence of graphs. 
The inverse limit of the sequence with Hausdorff reflection gives a space homeomorphic to the original space.  
This idea was explored by Mardevsic (1960) \cite{mardevsic1960}, Smyth (1994) \cite{smyth1994inverse}, and Kopperman et al. (1997, 2003)\cite{Kopperman1997, Kopperman2003}. 

Asymptotic finite approximations of continuous mappings between such spaces are also important. They were investigated in  Charalambous (1991) \cite{charalambous1991approximate}, and 
Mardevsic (1993) \cite{mardevsic1993approximate}. 
In Debski and Tymchatyn (2017) \cite{debski2017cell} and (2018) \cite{debski2018cell}, there is a comprehensive (and up to date) approach to discrete approximations of complete metrizable topological spaces and continuous mappings. 

The above strictly topological approach, based only on open coverings and corresponding graphs and nerves, has limitations if geometrical aspects (like geodesics and curvature) are considered.    

We are interested in asymptotic finite approximations of geometrical spaces. The geometry is to be understood here in the spirit of Busemann's (1955) book {\em The geometry of geodesics} \cite{Busemann1955}. His approach is based on axioms that characterize the geometry of a class of metric spaces (called G-spaces or geodesic metric spaces), and may be summarized as follows.  The spaces are metric and finitely compact, i.e any bounded infinite sequence of points has at least one limit point.  Any two points can be connected by a geodesic. 

The notion of Riemannian manifold itself is abstract, complex  and involves sophisticated notions of atlas and metric tensor; it is hard to represent them by limits of finite structures.

Summarizing the short review of the state of the art, constructive (and computational) approach to geometry was a challenge for a long time. It was considered in an abstract topological setting, or based on the complex notion of Riemannian manifold. 

Perhaps the right way to solve the problem is to find a generic method for constructing geometrical spaces as limits of finite structures instead of approximating already defined ones, that are, in fact, complex abstract notions. 


\subsection{A sketch of our approach }

We are going to explore the following idea. 
A compact geodesic metric space, say $(X, d)$, can be tessellated (dividing into uniform and regular closed subsets) instead of being covered by open sets. 

The adjacency, in a tessellation, can be represented as a graph, say $G_1$. The tessellation can be refined into a next tessellation (say $G_2$) in the way that each element (vertex) of $G_1$ is subdivided uniformly according to a fixed inductive pattern. Continuing this step by step, we get the inverse sequence of the graphs $(G_1, G_2, G_3, \dots )$. Usually, the inverse limit of the sequence (say $L$) has the Cantor topology. Each point of $L$ is the limit of converging nested closed subsets of $X$ that corresponds to one point of $X$. However, many points of $L$ may correspond to the same point of $X$. 

The idea of this paper is to define (without using the original metric $d$) a metric (say $d_n$), for each graph $G_n$,  such that the limit of the sequence $(d_1, d_2, d_3, \dots )$ gives the original metric $d$ of the space $X$.  

The idea of tessellation may be abstracted to the following one. 

Graph $G_k$ approximates geometrical space such that vertices corresponds to points (in the space) that uniformly span the space, i.e. the edge distance of any two neighboring vertices are one and the same (converging to $0$ as $k \to \infty$), and  for each point of the space there is a vertex with the distance to that point less than the edge distance. 

If the pattern of the construction of the sequence $(G_1, G_2, G_3, \dots )$ is {\em sufficiently regular}, then there is also a geometrical structure encoded in the form of a pre-metric for any of the graphs, say $p_n$ for graph $G_n$. Pre-metric $p_n$ means that only distance between two adjacent vertices (connected by edge in the graph $G_n$) is determined. It is reasonable to assume that this distance is constant and minimal for a fixed $n$.  

Pre-metric $p_n$ can be extended to a metric $d_n$ on the whole graph $G_n$. It can be done in many uniform (for all graphs in the sequence) ways. 
For such metric $d_n$, the distance between two neighboring vertices is minimal, i.e. the distance between two not adjacent vertices is always bigger than the minimal one. 

The necessary constrain for this extension is that the sequence of such metrics $d_n$ converges to a pseudo-metric (say $d'$) on the space $L$. Pseudo-metric means that distance between two different points may be not be equal $0$.  The quotient $L/_{d'}$ (where points, with distance $0$ between them, are identified) is a metric space with the metric denoted by $\bar{d}$.   

{\em Is $(L/_{d'}, \bar{d})$  isomorphic (distance preserving) to the original space $(X, d)$?}

If the answer is YES, then the sequence $((G_n, d_n); n\in N )$, where  $N$ denotes the set of natural numbers without zero, may be considered as finite approximations of the original metric space $(X,d)$. 

Even if the answer is NO, then the metric space  $(L/_{d'}, \bar{d})$ may be of some interest.  


A simple and natural extension of the pre-metric $p_n$ is to define a metric as the length of shortest (according to $p_n$) paths in the graph $G_n$. Usually, it gives Taxicab-like metrics. 
Then, in the limit $(L/_{d'}, \bar{d})$, for any two points,  there may be infinitely many geodesics between them. 
Most of the {\em important} geometries have the property that  geodesics are unique locally. 
Hence, the extensions of $p_n$ must be more smart in order to restore the original metric $d$ on the basis of the sequence $((G_n, d_n); n\in N )$ alone. 

The adjacency, expressed in the form of graphs, stores mainly the topological information on the metric space $(X, d)$. However, metric is much more reach structure than topology. Usually, with some effort, a sequence $((G_n, d_n); n\in N )$  can be constructed such that it approximates the original metric space $(X, d)$. This may give rise to define geometrical notions like geodesics in $(G_n, d_n)$ that  approximate geodesics in $(X, d)$. 

Then, the sequence of graphs (with metric and geodesics) may be seen as finite approximations of the original geometrical space $(X, d)$. Straight lines and curves, plains and surfaces, and their higher dimensional analogs can be finitely approximated in $(G_n, d_n)$ where the parameter $n$ corresponds to the resolution scaling factor.  In computer graphics, each of these approximations is a subset of vertices (pixels (2D) or voxels (3D)) that, along with their adjacency, form a subparagraph that may be considered as an approximation of a line or a plane, a figure on the plane, or a 3D manifold in $(X, d)$. 

Besides evident applications in computer graphics, the sequence $((G_n, d_n); n\in N )$ of finite and asymptotic approximations of the original geometrical space $(X, d)$ may be seen as the inductive construction of the original space. Then, the space itself is the abstraction (limit) of the inductive construction. Actually, the points of $X$ are abstractions of infinite sequences of vertices from finite graphs. 

If we forget the original space $(X, d)$, and consider only the sequence $((G_n, d_n); n\in N )$ as arbitrary sequence (defined without $(X, d)$) satisfying only some necessary conditions, then the resulting geodesic metric space $(L/_{d'}, \bar{d})$ is a combinatorial construction of an intrinsic geometry, possible a novel one. 

This gives rise to state the hypothesis that any compact geodesic metric space (including compact Riemannian manifolds) can be constructed in this very combinatorial way. 

Geodesics are crucial in geometry. 
A shortest path in graph $G_n$ (where the edge length is determined by $p_n$) is not necessary a geodesic according to metric $d_n$. It is for Taxicab metrics. Hence, finite geodesics in the graphs must be given a special definition; it is done in Section \ref{n-geodesic}. 


Then, {\em geodesic} in the metric space $(L/_{d'}, \bar{d})$ can be defined (see Section \ref{geodesic} as a uniform limit (if it exists) of finite geodesics of the graphs. This simple concept of geodesic is sufficient to derive sophisticated geometrical notions like curvature that is explored in Section \ref{main-curvature}.  

All topological features (like homotopy type, covering dimension and embedding dimension), and the geometrical features (like flatness, elliptic or hyperbolic curvature) of the space $(L/_{d'}, \bar{d})$ can be deduced from the pattern of inductive  construction of the sequence $((G_n, d_n); n\in N )$ of finite geometrical structures (graphs) with metrics. 

Note that geometrical spaces, constructed in this inductive and combinatorial way, are stand-alone constructions without an ambient space, and present the intrinsic point of view in geometry.  
The crucial feature of the proposed approach is its computational aspect, i.e.  the geometries are (by their constructions) uniformly approximated by finite graphs with metrics.


\section{Intrinsic versions of classic geometries }
\label{examples}

\begin{figure}
\centering
	\includegraphics[width=0.7\textwidth]{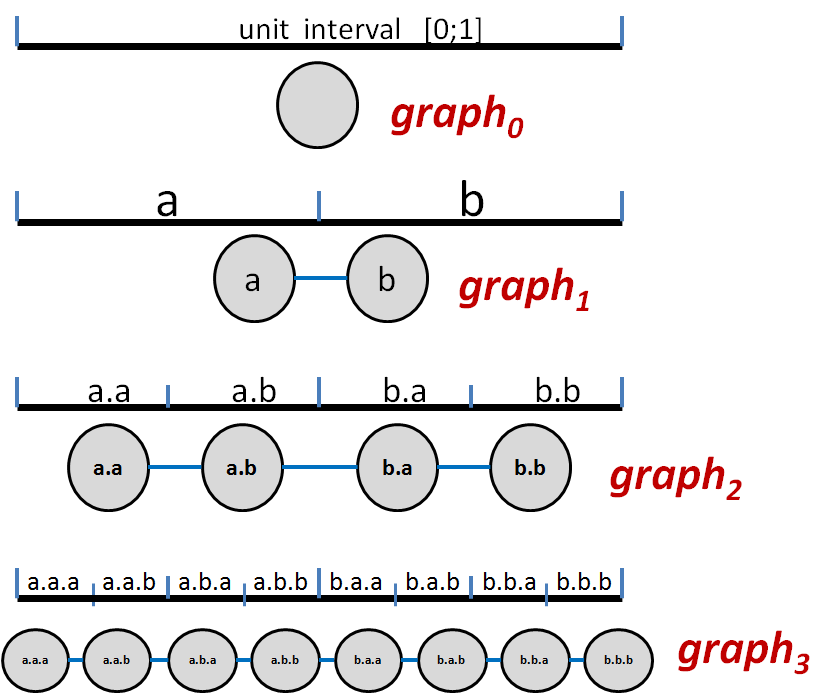}
	\caption{The first four consecutive graphs of the unit interval} 
	\label{unit-interval}
\end{figure}
\begin{figure}
\centering
	\includegraphics[width=0.7\textwidth]{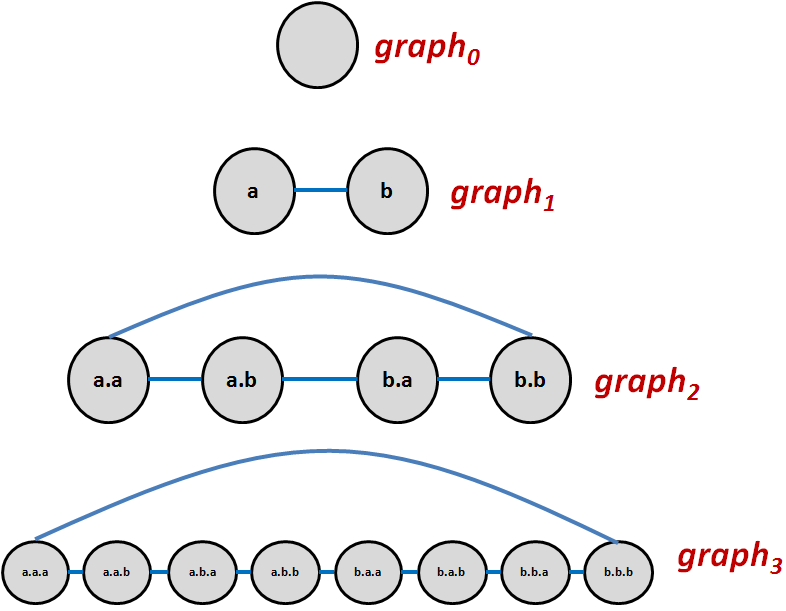}
	\caption{The first four consecutive graphs of the flat circle } 
	\label{circle}
\end{figure}
\begin{figure}
\centering
	\includegraphics[width=0.9\textwidth]{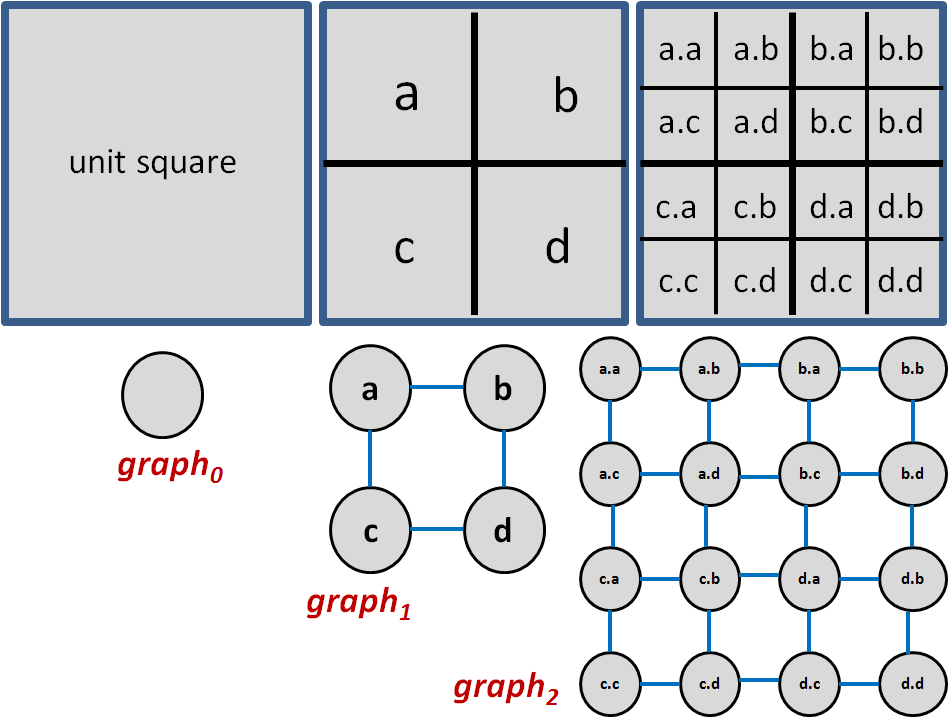}
	\caption{The first three quadrilateral tessellations of the unit square} 
	\label{unit-square}
\end{figure}

Let us start with the simple intrinsic geometry of the unit interval $[0, 1] \subset \mathbb{R}$. It can be tessellated consecutively in the way shown in Fig. \ref{unit-interval}. Following the notation from our previous paper on the grounding for Continuum \cite{C}, let the tessellation pattern be denoted $\mathbb{E}^1$. Let the resulting sequence of graphs be denoted 
$(G^{\mathbb{E}^1}_k; \ n\in N )$. 
 
The metric $d^{\mathbb{E}^1}_n$ can be defined in the following natural way. For any vertices $x$ and $y$ in $G^{\mathbb{E}^1}_n$, the distance $d^{\mathbb{E}^1}_n (x,y)$ is the length of the path between them divided by the diameter of the graph $G^{\mathbb{E}^1}_n$. 

If the two endpoints of the unit interval are merged, then the consecutive tessellations pattern (denoted  $\mathbb{S}^1$) gives the sequence of graphs $(G^{\mathbb{S}^1}_n; \ n\in N )$, see Fig. \ref{circle}. The distance $d^{\mathbb{S}^1}_n (x,y)$ is defined as the length of the shortest path between them divided by the diameter of the graph $G^{\mathbb{S}^1}_n$. 

The sequence $((G^{\mathbb{S}^1}_n, d^{\mathbb{S}^1}_n); \ n\in N )$ determines the intrinsic geometry of the well known {\em flat circle}. 



The next example concerns the unit square, i.e. the set $[0,\ 1] \times [0,\ 1] \subset \mathbb{R}^2$.  The square can be divided into four the same small squares. Each of the small squares can be divided into the next smaller squares, and so on. The sequence of 
graphs corresponding to the consecutive tessellations (see Fig. \ref{unit-square} is denoted $(G^{\mathbb{E}^2}_n; \ n\in N )$. 

For any vertices $x$ and $y$ of $G^{\mathbb{E}^2}_n$,  the natural  metric $d^{taxi}_n (x,y)$ is the length of the shortest path between them divided by $n$. Actually, it is the Taxicab metric. The limit of these metrics is the metric on the unit square determined by the norm $l_1$. For $(G^{\mathbb{E}^2}_n, d^{taxi}_n)$, there are many geodesics between two distinct vertices. In the limit, for any two points, there are infinitely many geodesics. 

There are also different metrics. One of them is the Euclidean metric. 
We are going to construct the sequence $((G^{\mathbb{E}^2}_n, d^{\mathbb{E}^2}_n);\ n\in N )$ that defines (in the limit) the Euclidean geometry on the unit square $[0,\ 1] \times [0,\ 1] \subset \mathbb{R}^2$.  

It seems that the notion of lines is crucial for that metric. Let three vertices $x$, $y$, and $z$ (in  $G^{\mathbb{E}^2}_n$) be such that $x$ and $y$ are connected by edge in the graph, and $y$ and $z$ are connected. These three vertices lay in a line, if for any face of the graph, if $x$ and $y$ belong to this face, then $z$ does not belong to this face. This may be considered as a definition of three points lying on {\em line} in $G^{\mathbb{E}^2}_n$.  

For any two distinct vertices $x$ and $y$ there are exactly two vertices, say $z_1$ and $z_2$ such that there is line connecting $x$ and $z_i$,  and  there is line connecting $y$ and $z_i$. The Euclidean distance between $x$ and $y$, denoted $d_n^{\mathbb{E}^2}(x, y)$, is defined as $\sqrt{(d^{taxi}_n(x, z_1))^2 + (d^{taxi}_n(y, z_1))^2 } = \sqrt{(d^{taxi}_n(x, z_2))^2 + (d^{taxi}_n(y, z_2))^2 }$. 

The Euclidean distance is an  example (but the most important one) of metrics that can be defined on $G^{\mathbb{E}^2}_n$. 

Let us briefly characterize the class of such metrics.  
Any face in graph $G^{\mathbb{E}^2}_n$ is a quadrilateral. Let any of its edge be of length $1$.  Let the distance between the opposite vertices be defined as $o$. In order to satisfy the triangle inequality, $o \leq 2$. 
On the other hand, if all minimal distances are assumed to be  those and only those that correspond to the adjacency (edges) in the graph, then $o > 1$.   
 

\subsection{Geodesics in finite metric space $(G_n, d_n)$}
\label{n-geodesic}
A geodesic between two points means the shortest path between them such that its length is equal to the distance between these two points.  

Length of a shortest path between two vertices in graph $G_n$ is not necessary equal to the distance between the two vertices. The above finite metric space  $(G^{\mathbb{E}^2}_n, d^{\mathbb{E}^2}_n)$, related to the unit square, may serve as a good example, and as a source of intuitions. 

The definition of {\em $(G_n, d_n)$-geodesic}, introduced below, is general and tries to capture the idea of geodesic in such finite metric spaces. Note that we are interested in the limits of such finite spaces as $n \to \infty$. So that a $(G_n, d_n)$-geodesic may be seen as an approximation of a geodesic in the limit space; see formal definition in Section \ref{geodesic}. 

Let us recall that a path between two distinct vertices $x$ and $y$ in a graph is a sequence $(z_1, z_2, \dots z_k)$ of vertices such that $x=z_1$ and $y=z_k$, and there is edge between $z_i$ and $z_{i+1}$ for $1\leq i < k$. The length of a path is the sum of lengths of its edges.  
A shortest path in a graph is not necessary a geodesic. 

A {\em $(G_n, d_n)$-geodesic} between two vertices $x$ and $y$ in finite metric space $(G_n, d_n)$ is defined as a shortest path $(z_1, z_2, \dots z_k)$ between them that satisfies the following locally minimizing condition. 
%
\begin{itemize} 
\item 
For any $i=2,3, \dots , k-1$, 
\begin{itemize}
\item  
for all $z$ that are adjacent to $z_{i-1}$ and $z_{i+1}$: \\ 
$d_n(x, z_{i}) + d_n(z_{i}, y) \leq d_n(x,z) + d_n(z,y)$.  
\end{itemize} 
\end{itemize}

Note that the following condition is essential here.   

{\em 
$d_n(x,y)$ is minimal if and only if there is edge between $x$ and $y$. }

Hence, the vertices of graph $G_n$ may be viewed as points that uniformly span the space, the space that is abstraction of $G_n$ as $n \to \infty$. For any point of the abstract space, its minimal distance to vertices is less than the edge distance. 




All geodesics in the unit square are limits of $(G^{\mathbb{E}^2}_n, d^{\mathbb{E}^2}_n)$-geodesics as $n \to \infty$. 

A general definition of geodesics in the limit abstract spaces is proposed in Section \ref{geodesic}. 



\subsection{Intrinsic geometry of flat torus } 
\label{working-example}

The ``square'' flat torus is the quotient  $\mathbb{R}^2/_{\mathbb{Z}^2}$, where $\mathbb{Z}$ is the set of integers, and a group under the operation of addition.
It can be interpreted as the Cartesian (Euclidean) plane $\mathbb{R}^2$ under the identifications $(x, y) \sim (x + 1, y) \sim (x, y + 1)$.   The metric is inherited from $\mathbb{R}^2$ via the identification. Actually, it is a complex abstraction that involve the set of real numbers and classes of abstraction determined by the equivalence relation $\sim$.  Although, it is a common and ubiquitous way of defining new notions in Mathematics, we are going to show that this ``square'' flat torus and ``square'' flat Klein bottle, as geometric spaces, can be constructed as sequences, of finite graphs with metrics, of the from  $(G_n, d_n); \ n\in N)$.


 The construction is based on the following idea. 

A classic torus, embedded in Euclidean space $\mathbb{R}^3$, can be tessellated by quadrilaterals, however, not by flat ones  because of the curvature. It is possible to tessellate torus with 4 quadrilaterals. 

The next finer tessellation is done by dividing each of the 4 quadrilaterals into 4 smaller quadrilaterals. Then, each of the smaller quadrilaterals may be divided into 4 quadrilaterals, and so on. 
\begin{figure}
\centering
	\includegraphics[width=0.7\textwidth]{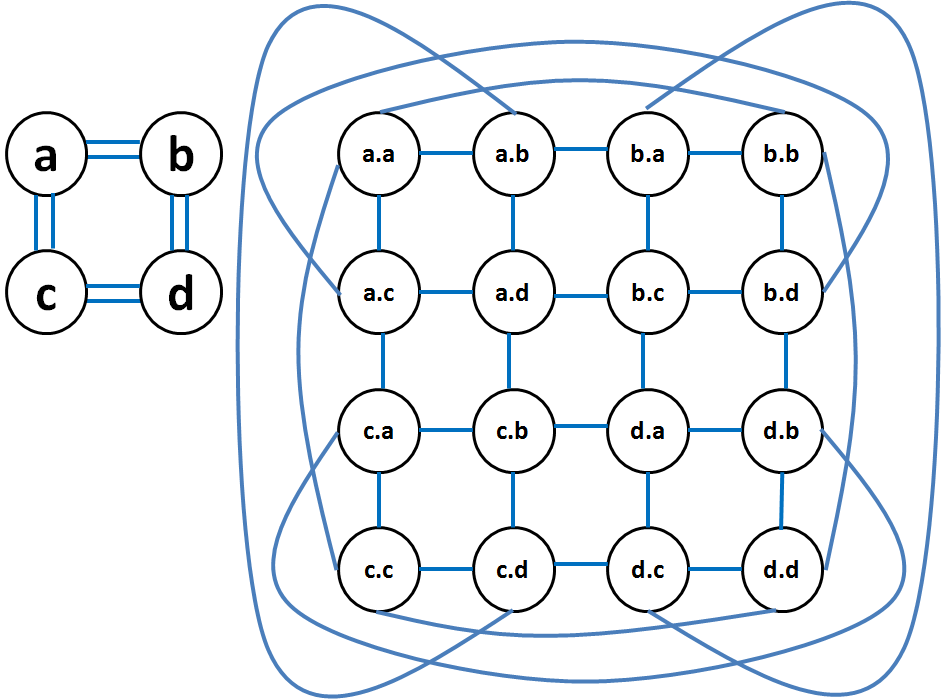}
	\caption{The duals to the first two quadrilateral tessellation graphs of torus} 
	\label{torus}
\end{figure}

Actually, any tessellation net (consisting of edges and vertices) is a graph. The dual (to this tessellation graph) is the graph where the vertices correspond to quadrilaterals, and edges correspond to the adjacency of the quadrilaterals. 

The first two dual graphs are shown in Fig. \ref{torus}. 
The dual graph of the first tessellation is a multi-graph. i.e.  the edges are duplicated. 

We may construct the infinite sequence of finer and finer tessellations and corresponding dual graphs. Let the sequence be denoted $(G^\mathbb{T}_n; \ n\in N )$. 

Now, let us abstract from a classic torus, embedded in $\mathbb{R}^3$, and consider only the sequence $(G^\mathbb{T}_n; \ n\in N )$ without any reference to that classic torus. 

It is clear, from Fig. \ref{torus}, that  Euclidean metric  $d^{\mathbb{E}^2}_n$ defined on $G^{\mathbb{E}^2}_n$ can be adjusted to $G^{\mathbb{T}}_n$ resulting in the metric denoted  $d^{\mathbb{T}}_n$. 

So that, the sequence $((G^{\mathbb{T}}_n, d^{\mathbb{T}}_n); \ n\in N)$ defines, in the limit, the geometry of the  ``square'' flat torus, see Section \ref{finite} for details. 

The flat torus is homeomorphic to an Euclidean torus. By the famous Nash embedding theorem, a flat torus can be isometrically embedded in $\mathbb{R}^3$ with the class 1 smooth mapping (see Borrelli et al. (2012) \cite{borrelli2012flat}) but not with the class 2. 

\begin{figure}
\centering
	\includegraphics[width=0.3\textwidth]{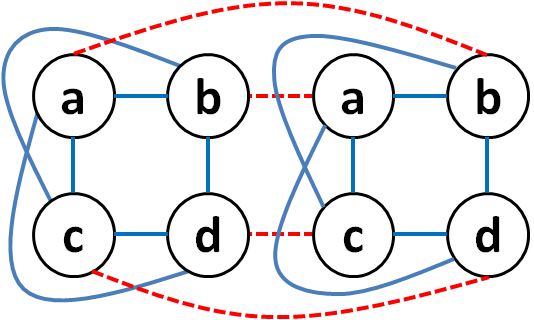}
	\caption{The dual of the first quadrilateral tessellation graph of Klein-bottle} 
	\label{Klein-bottle1}
\end{figure}
\begin{figure}
\centering
	\includegraphics[width=1.0\textwidth]{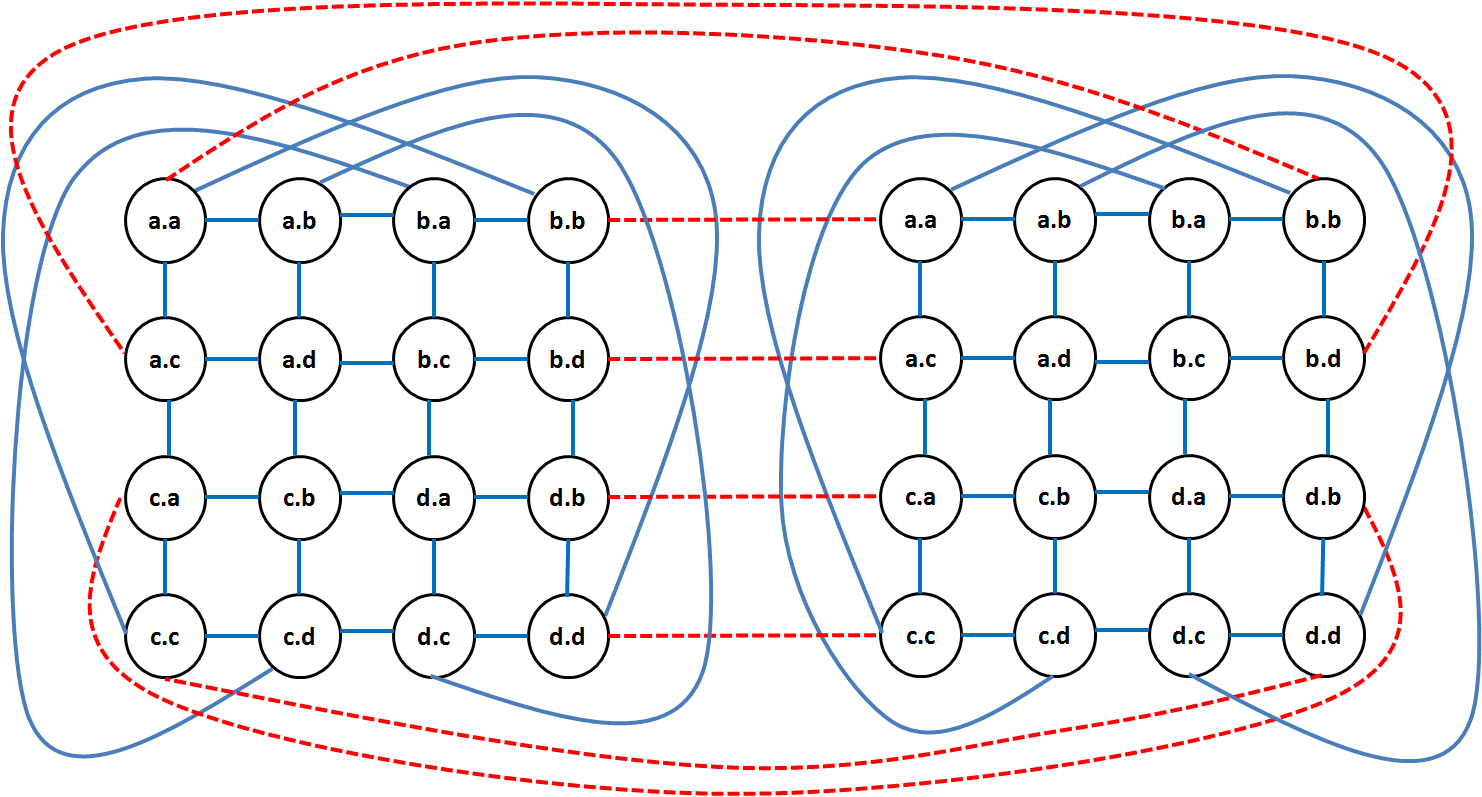}
	\caption{The dual of the second quadrilateral tessellation graph of Klein-bottle. It is composed of the duals of two M\"obius strips tessellation graphs (where edges are solid lines) by joining (by edges shown as dashed red lines) the borders of the strips} 
	\label{Klein-bottle2}
\end{figure}

The same idea can be applied to the ``square'' flat Klein bottle. A classic Klein bottle, embedded in $\mathbb{R}^4$,  can be tessellated by quadrilaterals. The simple argument for that is that the bottle can be constructed by joining the borders of two M\"obius strips, see Fig. \ref{Klein-bottle1} and Fig. \ref{Klein-bottle2}. 

So that, the infinite sequence $((G^{\mathbb{K}}_n, d^{\mathbb{K}}_n)); \ n\in N )$  of graphs (dual to finer and finer tessellations) and metrics derived from Euclidean metrics, can be constructed. The sequence determines (in the limit) the geometry of the ``square'' flat Klein bottle. Although it cannot be embedded in $\mathbb{R}^3$ (as a topological space), it is a concrete geometrical space that can be approximated by finite graphs with metrics. 

All geometrical features of flat torus and flat Klein bottle are encoded in the construction patterns of the corresponding sequences of graphs. 

To summarize this section, let us again emphasize the main idea of the paper  that a sequence of the form ($(G_n, d_n);\ n\in N$) constructed according to {\em a regular and uniform pattern} may be considered (in the limit) as an intrinsic geometrical space.


In the next subsection \ref{formal}, the above, a bit informal,  considerations are presented rigorously.

\subsection{Topology}
\label{formal}

First, let us consider the topological space resulting (in the limit) from the sequence $(G^\mathbb{T}_n; n\in N)$. Following the notation  from \cite{C},  $\mathbb{T}$ is called the pattern of construction of the sequence, and also  denotes the topological space. 

It is convenient to use the {\em dot} notation (ASN.1) to denote the sub-quadrilaterals resulting from consecutive divisions. The quadrilaterals (vertices) of $G^\mathbb{T}_1$, see left part of Fig.  \ref{torus}, are denoted by the letters $a,b,c$ and $d$. The quadrilaterals (vertices) of $G^\mathbb{T}_2$ are denoted by $t_1.t_2$, where $t_1$ and $t_2$ belong to the set  $\{ a; b; c; d \}$.  

The next finer tessellation results in quadrilaterals (vertices in $G^\mathbb{T}_3$) denoted by labels of the form $t_1.t_2.t_3$, where $t_1$, $t_2$ and $t_3$ belong to the set  $\{ a; b; c; d \}$. 

In general case, the vertices in the graph $G^\mathbb{T}_k$ are denoted by finite sequences of the form  $x=t_1.t_2. \dots  t_{k-1}.t_k$. 
Let $C^\mathbb{T}_k$ denote the set of all such $x$ of length $k$.  So that, $C^\mathbb{T}_k$ is defined as the set of vertices of the graph $G^\mathbb{T}_k$. The edges of $G^\mathbb{T}_k$ are determined by the adjacency relation between the quadrilaterals. 

For our purpose it is convenient to consider the adjacency relation (denoted $Adj^\mathbb{T}_{k}$) instead of the set of edges of the graph $G^\mathbb{T}_k$. 

For $i$ less or equal to the length of $x$, let $x(i)$ denote the prefix (initial segment) of $x$ of length $i$. 

Note that for any $x$ of length $k$, the sequence ($x(1), x(2),\ \dots \ x(k-1), x(k)$) may be interpreted as a sequence of nested quadrilaterals  converging to a point on the torus if $k$ goes to infinity. 

The relation $Adj^\mathbb{T}_{k}$ is symmetric, i.e. for any $x$ and $y$ in $C_k$, if $Adj^\mathbb{T}_{k}(x,y)$, then $Adj^\mathbb{T}_{k}(y,x)$. It is also convenient to assume that $Adj^\mathbb{T}_{k}$ is reflexive, i.e. any $x$ is adjacent to itself, formally $Adj^\mathbb{T}_{k}(x,x)$. 

Let $C^\mathbb{T}$ denote the union of the sets $C^\mathbb{T}_k$ for $k\in N$. 
The relations $Adj^\mathbb{T}_{k}$ (for $k\in N$) can be extended to $C^\mathbb{T}$, i.e. to the relation $Adj^\mathbb{T}$, in the following way. For any $x$ and $y$ of different length (say $n$ and $k$ respectively, and  $n<k$), relations $Adj^\mathbb{T}(x,y)$ and $Adj^\mathbb{T}(y,x)$ hold if there is $x'$ of length $k$ such that $x$ is a prefix (initial segment) of $x'$,  and $Adj^\mathbb{T}_{k}(x', y)$ holds, i.e.  $x'$ is of the same length as $y$ and is adjacent to $y$. 

Consider infinite sequences of the form  $( t_1.t_2. \dots  t_{k-1}.t_k. \ \dots )$. Let $C^{\mathbb{T}\infty}$ denote the set of such sequences. By the construction of ($G^{\mathbb{T}}_k; k\in N$), any such infinite sequence corresponds to a sequence of nested quadrilaterals (on an Euclidean 2--torus) converging to a point. However, there may be four  such different sequences that {\em converge} to the same point of the torus.  

For an infinite sequence, denoted by $u$, let $u(k)$ denote its initial finite sequence (prefix) of length $k$. 

The set $C^{\mathbb{T}\infty}$ may be considered as a topological space (Cantor space) with topology determined by the family of clopen sets $\{ U_x: x \in C \}$ such that $U_x := \{u: u(k) = x\}$ where $k$ is the length $x$. Note that the adjacency relation $Adj^{\mathbb{T}}$ is not used in the definition.  

There is another topological and geometrical structure on the set $C^{\mathbb{T}\infty}$ determined by $Adj^{\mathbb{T}}$. 

Two infinite sequences $u$ and $v$ are defined as {\em adjacent} if for any $k\in N$, the prefixes $u(k)$ and $v(k)$ are adjacent, i.e.  $Adj^{\mathbb{T}}(u(k) , v(k) )$ holds. Let this adjacency relation, defined on the infinite sequences, be denoted by $Adj^{\mathbb{T}\infty}$. Note that any infinite sequence is adjacent to itself. Two different adjacent infinite sequences may have a common prefix.  

The transitive closure of $Adj^{\mathbb{T}\infty}$ is an equivalence relation denoted $\sim$. Let the quotient set be denoted by $C^{\mathbb{T}\infty}/_{\sim}$, whereas its elements, i.e. the equivalence classes, be denoted by $[v]$ and $[u]$. Actually, $C^{\mathbb{T}\infty}$ is known as the inverse limit of $(G^{\mathbb{T}}_k; k\in N)$, whereas $C^{\mathbb{T}\infty}/_{\sim}$ as the Hausdorff reflection of the limit relatively to the equivalence relation $\sim$. 

There are rational and irrational points (equivalence classes) in $C^{\mathbb{T}\infty}/_{\sim}$. For the flat torus, each rational equivalence class has four elements, whereas the irrational classes are singletons. Each rational class corresponds to a vertex of a tessellation graph, and equivalently to a face in graph $G^{\mathbb{T}}_k$ for some $k$. 

By the construction of $G^{\mathbb{T}}_k$, any point of the initial Euclidean 2-torus corresponds exactly to one equivalence class (a point in  $C^{\mathbb{T}\infty}/_{\sim}$) and vice versa. 

The relation $Adj^{\mathbb{T}\infty}$ determines natural topology on the quotient set $C^{\mathbb{T}\infty}/_{\sim}$.   

Usually, a topology on a set is defined by a family of open subsets that is closed under finite intersections and arbitrary unions. The set and the empty set belong to that family.  Equivalently (in the Kuratowski style), the topology is determined by a family of closed sets; where finite unions and arbitrary intersections belong to this family. 

Let us define the base for the closed sets as the collection of the following neighborhoods of the points of $C^{\mathbb{T}\infty}/_{\sim}$. For any equivalence class $[v]$, i.e. a point in $C^{\mathbb{T}\infty}/_{\sim}$, the neighborhoods (indexed by $k\in N$) are defined as follows. 

$$
U^{[v]}_k := \{ [u] \in C^{\mathbb{T}\infty}/_{\sim}: \  \exists_{v'\in [v]} \exists_{u'\in [u]}\   Adj^{\mathbb{T}}(u'(k), \ v'(k)) \}
$$

 Note that $u'(k)$ may be equal to $v'(k)$. 

A sequence $(e_1, \ e_2,\ \dots \ e_n, \ \dots \ )$ of  equivalence classes of $\sim$ (elements of the set$C^{\mathbb{T}\infty}/_{\sim}$) converges to $e$, if for any $i$ there is $j$ such that for all $k>j$, \ \ \  $e_k \in U^e_i$. 

The set of rational classes is dense in $C^{\mathbb{T}\infty}/_{\sim}$. 

The topological space $C^{\mathbb{T}\infty}/_{\sim}$ is a Hausdorff compact space.  Let the space be denoted by $\mathbb{T}$. 
The sequence of graphs $(G^{\mathbb{T}}_n; n\in N)$ may be seen as finite approximations of the topological space $\mathbb{T}$, more exact if $n$ is grater. 

The space $\mathbb{T}$ is homeomorphic to 2-torus embedded in the  Euclidean space $\mathbb{R}^3$.

\subsection{Geometry}

Flatness and curvature are geometrical notions. 
The sequence  $(G^\mathbb{T}_n; n\in N)$ contains a pre-geometric structure specific to the``square'' flat torus. 

For any $n$ there is a wide spectrum of metrics that can be defined on $G^{\mathbb{T}}_n$ preserving the property that the distance between adjacent vertices is the minimal one. The simplest metric is the Taxicab metrics that, from the geometric point of view, is not interesting. 

Let us choose the metric $d^{\mathbb{T}}_n$ (defined in Section \ref{working-example}) on $G^{\mathbb{T}}_n$. It is  derived from the Euclidean metric  $d^{\mathbb{E}^2}_n$ defined on $G^{\mathbb{E}^2}_n$.

Let us define function 
$\bar{d}^{\mathbb{T}}: C^{\mathbb{T}\infty} \rightarrow \mathbb{R}$ in the following way. 
$$
\bar{d}^{\mathbb{T}}(u,v) := \lim_{n \to \infty} d^{\mathbb{T}}_n(u(n),v(n))
$$ 
It is a pseudo-metric, i.e. a generalization of a metric where the distance between two distinct points may be zero.

For that particular case, it is clear, that $u \sim v$ if and only if $\bar{d}^{\mathbb{T}}(u,v) = 0$. 

So that, the metric, denoted $d^{\mathbb{T}}$, is defined on $C^{\mathbb{T}\infty}/_{\sim}$ as follows. 
$$
d^{\mathbb{T}}([u], [v]) := \bar{d}^{\mathbb{T}}(u,v)
$$

Let the metric space $(C^{\mathbb{T}\infty}/_{\sim}, d^{\mathbb{T}})$ be denoted $(\mathbb{T}, d^{\mathbb{T}})$. 

For any $n$, $(G^{\mathbb{T}}_n, d^{\mathbb{T}}_n)$-geodesics  in $G^\mathbb{T}_k$ are constructed  according to the definition in subsection \ref{n-geodesic}. 
 
Since the metric $d^{\mathbb{T}}_n$ inherits from the Euclidean metric $d^{\mathbb{E}^2}_n$, it is clear that any geodesic in $(\mathbb{T}, d^{\mathbb{T}})$ is a limit of $(G^{\mathbb{T}}_n, d^{\mathbb{T}}_n)$-geodesics.  

Hence, $(G^{\mathbb{T}}, d^{\mathbb{T}})$ is a geodesic metric space known as ``square'' flat torus. What does the flatness mean? What is specific in the construction pattern $\mathbb{T}$ that determines this very flatness and the curvature of the space in general? For the flat torus the answer is simple. Locally, the graphs $G^\mathbb{T}_n$ look the same as the  graphs for the unit square, see Fig. \ref{unit-square}. It is only an informal argument. The notion of curvature in $(G^{\mathbb{T}}, d^{\mathbb{T}})$ should be defined formally and must be based on the sequence $((G_n^{\mathbb{T}}, d_n^{\mathbb{T}}); n\in N)$ alone. It is done in Section \ref{main-curvature}. 

The sequence may be seen as finite approximations of the space $(\mathbb{T}, d^{\mathbb{T}})$, more exact if $n$ is grater. Actually, the sequence may be seen as a simple combinatorial construction of the ``square'' flat torus, contrary to its  abstract definition presented at the beginning of this section. Although the construction leads to {\em the same} abstract  geometrical space, the sequence $((G_n^{\mathbb{T}}, d_n^{\mathbb{T}}); n\in N)$ has much more rich structure than the resulting abstract geometrical space.

All geometrical features of the abstract space can be deduced from the pattern of inductive construction of the sequence. Moreover, the sequence may be a basis for constructing another interesting geometries.  

Note that the above definitions are for a concrete and specific inverse sequence corresponding to the flat torus. The general definitions are presented in the next Section \ref{finite}. 

The examples give rise to introduce a general notion of fully computational geodesic metric space where curvature can be defined in a nonstandard way.

\section{Generalization }
\label{finite}

We are going to generalize the notions introduced in the previous sections.  

Let ($G^{\mathbb{P}}_k,\phi^{\mathbb{P}}_{(k+1, k)}; k\in N$) be a sequence of graphs and mappings generated by an arbitrary pattern of inductive construction (denoted $\mathbb{P}$) such that: 
\begin{itemize}
\item
$C^{\mathbb{P}}_k$ denotes the set of vertices of $G^{\mathbb{P}}_k$. For $k\neq k'$, the sets $C^{\mathbb{P}}_k$ and $C^{\mathbb{P}}_{k'}$ are disjoint. 
\item  Mappings  $\phi^{\mathbb{P}}_{(k+1, k)}: C^{\mathbb{P}}_{k+1} \rightarrow C^{\mathbb{P}}_k$, for $k\in N$, are surjective, i.e. the image of the domain and the codomain are equal. 
\end{itemize}

Let $v$ denote an infinite sequence such that its $k$-th element belongs to $C^{\mathbb{P}}_k$. Let $v(i)$ denote $i$-th element of $v$. 

A {\em thread} is defined as an infinite sequence $v$ such that for any $k$, \ \   $v(k) = \phi^{\mathbb{P}}_{(k+1, k)}(v(k+1))$.  

Let $C^{\mathbb{P}\infty}$ denote the set of all threads. 

If for any $k$ and any vertex $x$ of $G^{\mathbb{P}}_k$, the number of threads $v$, such that $v(k) = x$, is infinite, then 
 the sequence ($G^{\mathbb{P}}_k,\phi^{\mathbb{P}}_{(k+1, k)}; k\in N$) is called an {\em inverse sequence}. 

Let $C^{\mathbb{P}}$ denote the union $\bigcup_{k\in N}C^{\mathbb{P}}_k$. 

Inverse sequence is a special case of inverse systems (see \cite{Kopperman2003}, \cite{smyth1994inverse}, and \cite{debski2018cell}) where topological spaces are substituted for finite graphs, and directed set is substituted for linear order. 

The inverse mappings determine a tree-like partial order on the set $C^{\mathbb{P}}$. 
  
For $x\in C^{\mathbb{P}}_k$ and $i<k$, let $x(i)$ denote the {\em ``prefix''} of $x$ of length $i$ defined as follows.  $x(k-1) := \phi^{\mathbb{P}}_{(k, k-1)}(x)$, \ \  $x(k-2) := \phi^{\mathbb{P}}_{(k-1, k-2)}(\phi^{\mathbb{P}}_{(k, k-1)}(x))$, and so on. 

Let the partial order {\em ``being a prefix of''} be denoted by $\leq_{\mathbb{P}}$. That is,  $x\leq_{\mathbb{P}}y$ iff $x$ is a prefix of $y$ or $x = y$. 

It is an improper use of the notion of prefix as an initial segment of a sequence. However, it is a convenient notation that operates on the tree-like partial order instead of the inverse mappings, and is similar to the notion of prefix from the preceding sections. 

We say that $x$ and $y$ are of the same length if they belong to the same $C^{\mathbb{P}}_k$. 

Let the adjacency relation $Adj^{\mathbb{P}}_k(x,y)$ denote that either   there is an edge connecting the vertices $x$ and $y$ in graph $G_k^{\mathbb{P}}$, or they are the same.  The extension of  $Adj^{\mathbb{P}}_k$ to $C^{\mathbb{P}}$, denoted $Adj^{\mathbb{P}}$, is defined in the very similar way as in  Section \ref{formal}. 

For a thread $v$, let $v(k)$, as an element of $C^{\mathbb{P}}_k$, be also called {\em the initial segment (``prefix'')} of length $k$ of the thread $v$. It is justified by the fact that for any $i < k$, \ \ $v(i)$ is uniquely determined by the inverse mappings. 

For these new (more abstract and general) notions, to have the intended meaning, some restrictions are necessary on the patterns of inductive construction of the inverse sequences.  


\subsection{Regular patterns } 

A pattern $\mathbb{P}$ is called {\em regular} if it satisfies the following two conditions. 
\begin{enumerate}
\item  {\em Connectedness.} 
If $x$ and $y$ (belonging to $C_k^{\mathbb{P}}$, for some $k$) are adjacent, then there are two different elements of $C_{k+1}^{\mathbb{P}}$ (say $x'$ and $y'$) such that $x=x'(k)$ and $y=y'(k)$, and $x'$ and $y'$ are adjacent. 
\item {\em Consistency.} If $x$ of length $k$, and $y$ of length $k'$ are adjacent, then for any $i<k$ and any $j<k'$: \ \   $x(i)$ and $y(j)$ are adjacent. 
\end{enumerate}

For a regular pattern $\mathbb{P}$, the adjacency relation $Adj^{\mathbb{P}\infty}$ for threads $u$ and $v$ (elements of  $C^{\mathbb{P}\infty}$) is defined as follows.

\centerline{$Adj^{\mathbb{P}\infty}(v,u)$ iff for any $k\in N$:  $Adj^{\mathbb{P}}(v(k),u(k))$}

 Its transitive closure $\sim$ is an equivalence relation. So that, the Hausdorff reflection of  $C^{\mathbb{P}\infty}$ relatively to $\sim$, i.e. $C^{\mathbb{P}\infty}/_{\sim}$, is a compact Hausdorff space, 
where the topology is determined by the collection of the  neighborhoods of the points of $C^{\mathbb{P}\infty}/_{\sim}$ defined in the similar way as in Section \ref{formal}. 

For the flat torus (Section \ref{formal}), the corresponding pseudo-metric $d^{\mathbb{T}}$ is concrete example of the notion of asymptotic pseudo-metric defined below.

\subsection{Asymptotic pseudo-metric } 
\label{Asymptotic pseudo-metric}

Let $\mathbb{P}$ be a regular pattern, and let for any $k\in N$,  $d_k$ be a metric that extends pre-metric on $G_k$. The pre-metric is a Taxicab-like metric, and is determined by the edges of the graph, so that the distance (according to this pre-metric) between two vertices is the length of the shortest path between them. The extension of the pre-metric must satisfy the condition that the distance $d_k$ between two vertices is minimal if and only if they are neighbors, i.e. there is edge between them in the graph.   

If for any threads $u$ and $v$ in $C^{\mathbb{P}\infty}$   
 the following limit  exists
$$
\lim_{k \to \infty} d_k(u(k),v(k))
$$
and the convergence is uniform,  then 
$$
 d(u,v) := \lim_{k \to \infty} d_k(u(k),v(k))
$$
is called {\em  asymptotic pseudo-metric}.  
It is clear that the triangle inequality is satisfied. 

Asymptotic pseudo-metric $d$ will be identified with the sequence $(d_k; k \in N)$. 

Hence, it is reasonable to say that the sequence $(G^{\mathbb{P}}_k,\phi^{\mathbb{P}}_{(k+1, k)}, d_k; \ k\in N)$ approximates the corresponding geometrical space $(C^{\mathbb{P}\infty}/_{\sim_{d}}, \bar{d} )$, where $\bar{d}$ is the metric determined by pseudo-metric $d$. Let the space be called an {\em asymptotic metric space} and denoted $(\mathbb{P}, d)$. 

Note that $(\mathbb{T}, d^{\mathbb{T}})$  is an asymptotic metric space. 

Again, in the spirit of Busemann's (1955) book {\em The geometry of geodesics} \cite{Busemann1955}, we aim at defining geodesics. 
Geodesic (in metric space) is a curve which is everywhere locally a distance minimizer. Extensions of geodesic are unique. 
On the basis of asymptotic pseudo-metric we are going to define geodesics in a constructive and fully computational way.

\subsection{Geodesics } 
\label{geodesic}


Let $(\mathbb{P}, d)$ be an asymptotic metric space. 

Let $(G_k,   \phi^{\mathbb{P}}_{(k+1,k)}, d_k; k \in N)$ be the corresponding sequence of graphs with metrics. 

Definition of $(G_k, d_k)${\em-geodesic} in finite metric space $(G_k, d_k)$ was introduced in the subsection \ref{n-geodesic}. 

Let us consider the graph $G_k$ for a fixed $k$.   
A path between vertices $x$ and $y$, in the graph, is defined in the usual way as a finite sequence of vertices $(z_1, z_2, \dots, z_n)$ such that $z_1 = x$ and $z_n = y$, and for any $i= 1,2, \dots n-1$, there is edge between $z_{i}$ and $z_{i+1}$. 

For any path $(z_1, z_2, \dots, z_n)$ its {\em contraction} is defined as the longest sub-path $(z_{n_1}, z_{n_2}, \dots, z_{n_l})$ such that for any $i= 1,2, \dots l-1$, the consecutive vertices $z_{n_i}$ and $z_{n_{i+1}}$ are different.  

A $(G_k, d_{k_2})$-geodesic $(y_1, y_2, \dots, y_{m_2})$ is called an {\em extension} of $(G_{k_1}, d_{k_1})$-geodesic  $(x_1, x_2, \dots, x_{m_1})$, where $k_1 < k_2$, if:  
\begin{itemize}
\item the contraction of path $(y_1(k_1), y_2(k_1), \dots, y_{m_2}(k_1))$ is the same as the path $(x_1, x_2, \dots, x_{m_1})$, where $y_i(k_1)$ is the prefix of $y_i$ of length $k_1$.  
\end{itemize}  

Let a sequence $(g_k; \ \ k\geq n)$, where $g_k$ is a $(G_k, d_k)$-geodesic, satisfies the following condition: 
\begin{itemize}
\item  for any $k\geq n$, \   $g_{k+1}$ is an extension of $g_k$.  
\end{itemize}
If such minimal number $n$ exists, then let it be denoted by  $n_\gamma$. 
Then, the sequence $(g_k; \ \ k\geq n_\gamma)$ determines (in the limit) a curve (denoted $\gamma$) in $(\mathbb{P}, d)$. Let $u$ and $v$ denote the end-points of the curve, and $z$ a point on the curve between the two end-points. 

If for any such $z$, $d(u,v) = d(u,z) +d(z,v)$, then let $\gamma$ 
be called a {\em $(\mathbb{P}, d)$-geodesic}. 

Uniqueness of extensions of geodesic depends on the metric $d$ that is constructed on the basis of metrics $(G_n, d_n)$ for $n\in N$. 

By the above definition, the sequence of the beginning-vertices of $g_k$, and the sequence of the end-vertices of $g_k$ (for $k \geq n_{\gamma}$) determine two threads that are, respectively, the beginning-point, and the end-point of the geodesic $\gamma$. 

Note, that $g_k$ may be viewed as a linear sub-graph  of $G^{\mathbb{P}}_k$. Let the sub-graph be denoted by $G_k^{\gamma}$. 

For $k\geq n_\gamma$, let $C^{\gamma}_k$ denote the set of vertices of graph $G^{\gamma}_k$. 

Inverse mapping $\phi^{\gamma}_{(k+1,k)}: C^{\gamma}_{k+1} \rightarrow C^{\gamma}_k$ is constructed in the following way. For any vertex $y$ in $g_{k+1}$ let 

\centerline{$\phi^{\gamma}_{(k+1,k)}(y) := x$ such that $y(k) = x$. }

Hence, the sequence $(G^{\gamma}_k, \phi^{\gamma}_{(k+1,k)}, d_k; \ k\geq n_\gamma)$ has been constructed that determines the intrinsic stand-alone geometric space corresponding to $\gamma$. Let the space  be denoted by $(C^{\gamma}/_{\sim_d}, \bar{d})$, and let it be identified with the geodesic $\gamma$. It is homeomorphic to the unit interval.

\subsection{Summary of the section} 

For an asymptotic metric space $(\mathbb{P}, d)$, we have defined a fully computational notion of geodesic. This gives rise to define the notion of {\em asymptotic geodesic metric space} if for any two points of the space there is a $(\mathbb{P}, d)$-geodesic between them. 

Note that $(\mathbb{T}, d^{\mathbb{T}})$, i.e flat square torus,   is an asymptotic geodesic metric space. 

It seems that asymptotic geodesic metric spaces may be considered as a computational and constructive version of G-spaces introduced by Busemann (1955) \cite{Busemann1955}. 

It is important to note that the extrinsic geometry of a $(\mathbb{P}, d)$-geodesic, in the ambient space $(\mathbb{P}, d)$, is the same as the intrinsic geometry of the geodesic itself, i.e. $(C^{\gamma}/_{\sim_d}, \bar{d})$. This gives rise to define a new notion of curvature that is different than the sectional curvature for Riemannian manifolds. It is done in the next sections.  

Examples of combinatorial constructions of intrinsic geometries in Section \ref{working-example} are  to support the view that the proposed approach is generic. 

Let us emphasize once again that $(\mathbb{P}, d)$ is an abstraction of the sequence $(G^\mathbb{P}_n,  \phi^{\mathbb{P}}_{(n+1,n)}, d_n; \ n \in N)$. The notions of  $(\mathbb{P}, d)$-geodesic and curvature (to be  introduced in the Section \ref{main-curvature}) are defined by means of finite structures of the sequence. 

Actually, these notions can not be defined in  $(\mathbb{P}, d)$ alone where the finite structures were abstracted to abstract space. This constructive and combinatorial approach (based on $(G^\mathbb{P}_n,  \phi^{\mathbb{P}}_{(n+1,n)}, d_n; \ n \in N)$)  is crucial, and seems to be novel in the Foundations of Geometry.

\section{Flat, hyperbolic and elliptic circles }   
\label{hyperbolic and elliptic circles} 

\begin{figure}[h]
	\centering
	\includegraphics[width=0.99\textwidth]{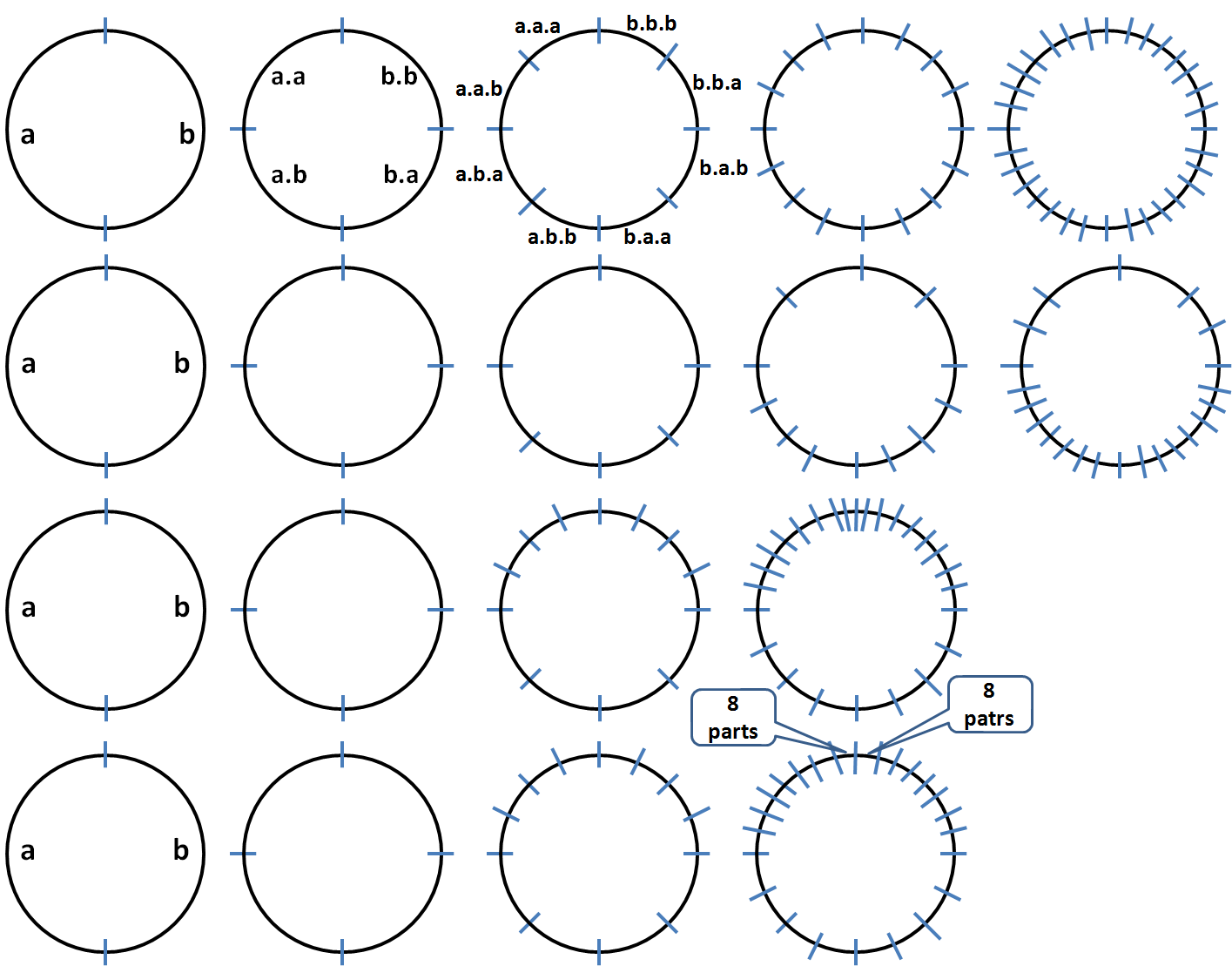}
	\caption{Rows from the top: The first row is for graphs of a flat circle. The second row is for graphs of a circle that is elliptic at one point. The third row is for graphs of a circle that is finitely hyperbolic at one point The fourth row is for graphs of a circle that is infinitely hyperbolic at one point }
	\label{circle-flat-eliptic-hiperbolic}
\end{figure}

Circle is one of the simplest geometrical spaces. It is of  elliptic curvature, and its common meaning is grounded in Euclidean plane. 
 
There is also a hyperbolic circle embodied in the hyperbolic plane $\mathbb{H}^2$, where the circumference of circle grows exponentially relatively to the radius. 

The unit interval $[0,\ 1] \subset \mathbb{R}$ can be represented as an intrinsic geometrical space in a simple way shown in Fig. \ref{unit-interval}. Circle can be made from the unit interval just by identifying the ends of the interval, see Fig. \ref{circle}. It is homeomorphic to an Euclidean circle embodied in $\mathbb{R}^2$. As an intrinsic geometrical space, it is the well known flat circle.  

A common view in Mathematics is that the only intrinsic geometry of circle is the flat one. Circle can be considered as a $1$-dimensional Riemannian manifold only. For this reason, sectional curvature cannot be defined for the circle. So that such manifold has no curvature at all. It is a continuum considered as actual infinity with the only structures of order and distance inherited from the unit interval where two end-points were merged. 

Our approach is different. A geometrical space (as continuum) is an abstraction of a sequence of the form  $(G_n,   \phi^{\mathbb{P}}_{(n+1,n)}, d_n; \ n \in N)$. Once this abstraction is acknowledged  as a fundamental idea in Geometry,  constructions of intrinsic geometries corresponding to hyperbolic circle and elliptic circle are possible, and are based on sequences of the form  $(G_n,   \phi^{\mathbb{P}}_{(n+1,n)}, d_n; \ n \in N)$. 



In order to grasp the intrinsic meaning of elliptic and hyperbolic curvatures of circles, we are going to introduce a new notion of curvature, different than the classic notion of sectional curvature in Riemannian manifolds. 

For an asymptotic geodesic metric space $(\mathbb{P}, d)$, determined by sequence $(G^\mathbb{P}_n,   \phi^{\mathbb{P}}_{(n+1,n)}, d_n; \ n \in N)$ a curvature at a point and direction determined by a geodesic, may be defined on the base of the following simple idea.  


%
{\em  Elliptic curvature} at a point means that if $k$ goes to infinity, then locally at the point, the length of a $(G_k^\mathbb{P}, d_k)$-geodesic ending at this point is getting shorter. This may be interpreted as gravitation force attracting to this point. 

{\em Hyperbolic curvature} at a point means that if $k$ goes to infinity, then locally at the point, the length of a $(G_k^\mathbb{P}, d_k)$-geodesic ending at this point is getting longer. This, in turn, may be interpreted as repulsive force from this point, and interpreted as anti-gravity or the electromagnetic repulsion. 



In order to explain the idea, let us consider the intrinsic  geometrical space corresponding to the flat circle, see Fig. \ref{circle-flat-eliptic-hiperbolic}, the first row from the top. Let its pattern be denoted by $\mathbb{O}$. Let the sequence ($G^{\mathbb{O}}_k, \phi^{\mathbb{O}}_{(k+1,k)}, d^{\mathbb{O}}_k;\ k\in N $) denote the corresponding inverse sequence with metric $d^{\mathbb{O}}_k(x,y)$ defined as the length of the shortest path between $x$ and $y$ in graph $G^{\mathbb{O}}_k$, (where the length of any edge is $1$), divided by the diameter of the graph.  

The graph $G^{\mathbb{O}}_1$ consists of two vertices $a$ and $b$. The set of vertices of graph $G^{\mathbb{O}}_2$ is $\{ a.a;\ a.b;\ b.a;\ b.b \}$. The inverse mappings are defined as follows. For any vertex $x$ of graph $G^{\mathbb{O}}_{k+1}$, \ \ $\phi^{\mathbb{O}}_{(k+1,k)}(x) := x(k)$. 
 
The space $(C^{\mathbb{O}\infty}/_{\sim_{d^{\mathbb{O}}}}, \bar{d^{\mathbb{O}}})$, also denoted $\mathbb{O}$, is homeomorphic and isometric to an Euclidean circle. 
However, it is not isomorphic (in the geometrical sense) because it is flat like  the interior of the unit interval, see Fig. \ref{unit-interval}. 

Let us choose the point that corresponds to the two threads $(b.b.b. \dots)$, where each element is $b$, denoted by $\bar{b}$, and $(a.a.a. \dots)$, where each element is $a$, denoted by $\bar{a}$. 

Note that $\bar{a}$ and $\bar{b}$ are adjacent, and the equivalence classes $[\bar{a}]$ and $[\bar{b}]$ are equal.  

Let $\bar{a}(k)$ and $\bar{b}(k)$ denote respectively the initial segment of $\bar{a}$ and $\bar{b}$ of length $k$.

\subsection{One-point elliptic curvature } 
\label{finite-elliptic}

Let us transform the inverse sequence ($G^{\mathbb{O}}_k, \phi^{\mathbb{O}}_{(k+1,k)}, d^{\mathbb{O}}_k;\ k\in N $) into the inverse sequence denoted  ($G^{e\mathbb{O}}_k, \phi^{e\mathbb{O}}_{(k+1,k)}, d^{e\mathbb{O}}_k;\ k\in N$), see Fig. \ref{circle-flat-eliptic-hiperbolic}, the second row from the top, 
in the following inductive way.  

Let $len^{\mathbb{O}}(x)$ denote the length of $x$, i.e. the number $k$ such that $x \in C^{\mathbb{O}}_k$, where $C^{\mathbb{O}}_k$ denotes the set of vertices of graph $G^{\mathbb{O}}_k$. Since the sets $C^{\mathbb{O}}_k$ and $C^{\mathbb{O}}_l$ are disjoint for  $k \neq l$, the length is well defined. 

For any $x$ of length $k$,  let $Suc(x)$ be the set of all $y\in C^{\mathbb{O}}_{k+1}$, such that $x = y(k)$. 

For any $k$ the set of verices $C^{e\mathbb{O}}_k$ of graph  $G^{e\mathbb{O}}_k$ is defined inductively as follows. 
 
\begin{itemize} 
\item
Let $C_1^{e\mathbb{O}} := C^{\mathbb{O}}_1$, and  $C_2^{e\mathbb{O}} := C^{\mathbb{O}}_2$. 
\item 
For $k=3$
$$
C_{3}^{e\mathbb{O}} := \{ a.a;\ b.b \} \cup \bigcup_{x\in C_{2}^{e\mathbb{O}}\ \& \ x\neq a.a\ \& \ x\neq b.b} Suc(x)) 
$$
\item 
For $k=4$
$$
C_{4}^{e\mathbb{O}} :=  \bigcup_{x\in C_{3}^{e\mathbb{O}}} Suc(x)) 
$$
\item 
For $k=5$
$$
C_{5}^{e\mathbb{O}} := \{ a.a.a;\ b.b.b \} \cup \bigcup_{x\in C_{4}^{e\mathbb{O}}\ \& \ x\neq a.a.a\ \& \ x\neq b.b.b} Suc(x)) 
$$
\item 
For $k=2n$
$$
C_{k}^{e\mathbb{O}} :=  \bigcup_{x\in C_{k-1}^{e\mathbb{O}}} Suc(x)) 
$$
and for any $y\in C_{k}^{e\mathbb{O}}$, let $\phi^{e\mathbb{O}}_{(k,k-1)}(y)$ be defined as $x\in C_{k-1}^{e\mathbb{O}}$ such that $y\in Suc(x)$. \\ 
Note that $\bar{a}(n+1)$ and $\bar{b}(n+1)$ (prefixes of  $\bar{a}$ and  $\bar{b}$ of length $n+1$)  belong to $
C_{k}^{e\mathbb{O}}$. 

\item
For $k=2n + 1$
$$
C_{k}^{e\mathbb{O}} := \{ \bar{a}(n+1);\ \bar{b}(n+1)\} \cup \bigcup_{x\in C_{k-1}^{e\mathbb{O}}\ \&\ x\neq \bar{a}(n+1)\ \&\ x\neq \bar{b}(n+1)} Suc(x)) 
$$

For any $y\in C_{k}^{e\mathbb{O}}$: \\ 
if $y =  \bar{a}(n+1)$ or $y=\bar{b}(n+1)$,  then  let $\phi^{e\mathbb{O}}_{(k,k-1)}(y):= y$. \\
Otherwise, let $\phi^{e\mathbb{O}}_{(k,k-1)}(y)$ be defined as $x\in C_{k-1}^{e\mathbb{O}}$ such that $y\in Suc(x)$.
\end{itemize}

Note that the length of elements of $C_{2n+1}^{e\mathbb{O}}$ that are prefixes of $\bar{a}$ or $\bar{b}$ is $n+1$, whereas the length of the most elements of $C_{2n+1}^{e\mathbb{O}}$ is equal or close to $2n+1$.  

Edges of graph $G^{e\mathbb{O}}_{k}$ are determined by adjacency relation $Adj^{\mathbb{O}}$ restricted to the set $C^{e\mathbb{O}}_k$ of the vertices of the graph. For the definition of the adjacency relation, see Sections \ref{formal} and \ref{finite}. 

For any $k$, let  $d_k^{e\mathbb{O}}$ denote metric on $C^{e\mathbb{O}}_k$ such that $d_k^{e\mathbb{O}}(x,y)$ is the length of the shortest path in the graph $G^{e\mathbb{O}}_{k}$, divided by the diameter of the graph.  The resulting  pseudo-metric is denoted by $d^{e\mathbb{O}}$.  The corresponding asymptotic geodesic metric space is denoted by $e\mathbb{O}$. 
 
 Note that now, the prefixes are taken relatively to the inverse mappings $\phi^{e\mathbb{O}}_{(k+1,k)}$, for $k \in N$. 
 
For any $u$ different than  $\bar{a}$ and $\bar{b}$, the distances $d_k^{e\mathbb{O}}(u(k),\bar{a(k)})$ and \\ $d_k^{e\mathbb{O}}(u(k),\bar{b}(k))$ are getting  shorter (if $k\to\infty$) comparing to other points, especially to $b.\bar{a}$ and $a.\bar{b}$ that are the antipodal points of $\bar{a}$ and $\bar{b}$ respectively. 

The  elliptic curvature at the point $[\bar{a}] = [\bar{b}]$ of the space $e\mathbb{O}$ seems to be equal $\frac{1}{2}$. The formal definition of curvature is introduced in Section \ref{curvature}.

The above method of construction of space $e\mathbb{O}$ is generic, i.e. it can be extended for any regular pattern. Also the number $2$ in $\frac{n+1}{2n + 1}$ (interpreted as the attraction parameter) may be bigger. Then, if approaching the point, the distance to that point is getting more shorter.

\subsection{One-point hyperbolic curvature } 
\label{one-point-hyperbolic}

Most models of hyperbolic space are constructed on the basis of Euclidean spaces, see the excellent exposition  by Cannon et al. (1997) \cite{Cannon1997}. We are going to present a generic method for constructing intrinsic and stand-alone hyperbolic spaces. 

On the base of the inverse sequence of the flat circle $\mathbb{O}$, two inverse sequences for two spaces are constructed below. One has a point of finite hyperbolic curvature, whereas the second one has a point of infinite hyperbolic curvature.

\subsubsection{Finite hyperbolic one-point curvature}  
\label{finite-hyperbolic}

Let $$C^{\mathbb{O}}:= \bigcup_{k\in N} C^{\mathbb{O}}_k$$

For a natural number ${\alpha}$, let us define function $h_{\alpha}: C^{\mathbb{O}} \rightarrow C^{\mathbb{O}}$ such that for any argument $x$ different than any prefix of $\bar{a}$ and different than any prefix of $\bar{b}$, \  $h_{\alpha}(x) := x$; and for any natural number $k$: \  
$h_{\alpha}(\bar{a}(k)) := \bar{a}(k+{\alpha})$ and 
$h_{\alpha}(\bar{b}(k)) := \bar{b}(k+{\alpha})$. 

Function $h_{\alpha}$ may be viewed as ${\alpha}$-weak repulsion force  at point $[\bar{a}]=[\bar{b}]$ of the circle $\mathbb{O}$.

For the clarity of presentation let us consider the case ${\alpha} = 2$. 

The function $h_2$ transforms the sequence 
$(G^{\mathbb{O}}_k, \phi^{\mathbb{O}}_{(k+1,k)}, d^{\mathbb{O}}_k;\ k\in N$) into the sequence $(G^{h_2\mathbb{O}}_k, \phi^{h_2\mathbb{O}}_{(k+1,k)}, d^{h_2\mathbb{O}}_k;\ k\in N)$, see Fig. \ref{circle-flat-eliptic-hiperbolic}, the second row from the bottom, in the following way. Here, $d^{h_2\mathbb{O}}_k$ is  the metric on $C^{h_2\mathbb{O}}_k$ such that $d^{h_2\mathbb{O}}_k(x,y)$ is defined as the length of the shortest path between vertices $x$ and $y$ of graph $G^{h_2\mathbb{O}}_k$, divided by the diameter of the graph. 

For any $x$ of length $k$ such that $h_2(x)=x$, let $Suc^{h_2}(x):= Suc(x)$, i.e. be defined as the set of all $y$, such that $x = \phi^{\mathbb{O}}_{(k+1,k)}(y)$. Otherwise, let $Suc^{h_2}(x)$ be the set of all $y$ of length $len^{\mathbb{O}}(h_2(x))$ such that $x<_{\mathbb{O}}y$, i.e. $x$ is a prefix of $y$. Function $len^{\mathbb{O}}$ is defined in subsection  \ref{finite-elliptic}. 
\begin{enumerate} 
\item
Let $C_1^{h_2\mathbb{O}} := C^{\mathbb{O}}_1$, and $C_2^{h_2\mathbb{O}} := C^{\mathbb{O}}_2$. 
\item
Suppose that $C_{k+1}^{h_2\mathbb{O}}$ and $\phi^{h_2\mathbb{O}}_{(k+1,k)}$ are  already defined. Then, let 
$$
C_{k+2}^{h_\mathbb{O}} := \bigcup_{x\in C_{k+1}^{h_2\mathbb{O}}} Suc^{h_2}(h_2(x)) 
$$
and  for any $y\in C_{k+2}^{h\mathbb{O}}$, let $\phi^{h_2\mathbb{O}}_{(k+2,k+1)}(y)$ be defined as $x\in C_{k+1}^{h_2\mathbb{O}}$ such that $y\in Suc^{h_2}(h_2(x))$. 

\end{enumerate}

Edges of graph $G^{h_2\mathbb{O}}_{k}$ are determined by adjacency relation $Adj^{\mathbb{O}}$ restricted to the set $C_{k}^{h_2\mathbb{O}}$ of the vertices of the graph.


The sequence $(G^{h_2\mathbb{O}}_k, \phi^{h_2\mathbb{O}}_{(k+1,k)}, d^{h_2\mathbb{O}}_k;\ k\in N)$ determines an asymptotic geodesic metric space denoted $h_2\mathbb{O}$. 

The distance $d_k^{h_2\mathbb{O}}$ from any vertex of $G^{h_2\mathbb{O}}_k$ to the vertices  of $G^{h_2\mathbb{O}}_k$ being prefixes of $\bar{a}$ or $\bar{b}$ is getting longer (if $k\to\infty$) comparing to other points, especially to the antipodal points of $\bar{a}$ and  $\bar{b}$. 

The finite hyperbolic curvature at the point $[\bar{a}] = [\bar{b}]$ of the space $h_2\mathbb{O}$ seems to be equal $2$. The formal definition of curvature is in Section \ref{curvature}. 

Note that the above method of construction of the space $h_2\mathbb{O}$ is generic, i.e. it can be extended for any regular pattern and any parameter $\alpha$.

\subsubsection{Infinite hyperbolic one-point curvature} 
\label{infinite-hyperbolic}
 
For a natural number ${\beta}$, let us define function $h^{\beta}: C^{\mathbb{O}} \rightarrow C^{\mathbb{O}}$ such that for any argument $x$ different than any prefix of $\bar{a}$ and any prefix of $\bar{b}$, \  $h^{\beta}(x) := x$; and for any natural number $k$: 
$h^{\beta}(\bar{a}(k)) := \bar{a}({\beta}k)$ and $h^{\beta}(\bar{b}(k)) := \bar{b}({\beta}k)$.  

The function $h^{\beta}$ may be viewed as ${\beta}$-strong repulsion at point $[\bar{a}]$ of the circle $\mathbb{O}$.  

Let us consider the case ${\beta} = 2$.  

The function $h^2$ transforms the inverse sequence ($G^{\mathbb{O}}_k, \phi^{\mathbb{O}}_{(k+1,k)}, d^{\mathbb{O}}_k;\ k\in N$) into the inverse sequence denoted by ($G^{h^2\mathbb{O}}_k, \phi^{h^2\mathbb{O}}_{(k+1,k)}, d^{h^2\mathbb{O}}_k;\ k\in N$), see Fig. \ref{circle-flat-eliptic-hiperbolic}, the first row from the bottom, in the following inductive way.   Here, $d^{h^2\mathbb{O}}_k$ is  the metric on graph $G^{h^2\mathbb{O}}_k$ such that $d^{h^2\mathbb{O}}_k(x,y)$ is the length of the shortest path between $x$ and $y$, divided by the diameter of the graph.  
  
For any $x$  such that $h^2(x)=x$, let $Suc^{h^2}(x) := Suc(x)$. Otherwise, let $Suc^{h^2}(x)$ be the set of all $y$ of length $len^{\mathbb{O}}(h^2(x))$ such that $x$ is a prefix of $y$.  
 
\begin{enumerate} 
\item
Let $C_1^{h^2\mathbb{O}} := C^{\mathbb{O}}_1$, and $C_2^{h^2\mathbb{O}} := C^{\mathbb{O}}_2$. 
\item
Suppose that $C_{k+1}^{h\mathbb{O}}$ and $\phi^{h^2\mathbb{O}}_{(k+1,k)}$ are  already defined. Then, let 
$$
C_{k+2}^{h^2\mathbb{O}} := \bigcup_{x\in C_{k+1}^{h\mathbb{O}}} Suc^{h^2}(h^2(x)) 
$$
and  for any $y\in C_{k+2}^{h^2\mathbb{O}}$, let $\phi^{h^2\mathbb{O}}_{(k+2,k+1)}(y)$ be defined as $x\in C_{k+1}^{h^2\mathbb{O}}$ such that $y\in Suc^{h^2}(h^2(x))$. 

\end{enumerate}
Edges of graph $G^{h^2\mathbb{O}}_{k}$ are determined by adjacency relation $Adj^{\mathbb{O}}$ restricted to the set $C_{k}^{h^2\mathbb{O}}$ of the vertices of the graph. 

The sequence $(G^{h^2\mathbb{O}}_k, \phi^{h^2\mathbb{O}}_{(k+1,k)}, d^{h^2\mathbb{O}}_k;\ k\in N)$ determines an asymptotic geodesic metric space denoted $h^2\mathbb{O}$. 

Intuitively, the hyperbolic curvature at point $[\bar{a}] = [\bar{b}]$ of the space $h^2\mathbb{O}$ (interpreted as the velocity of a particle approaching the point) is infinite. 

If, in the definition of function $h^2$, we substitute    $2^k$ for $2k$, then the normal pseudo-metric is extreme.  That is, the distance to $\bar{a}$ and to $\bar{b}$ from any other point is 1, i.e. the maximum. The distance between any two points (both are different than $\bar{a}$ and  $\bar{b}$) is $0$.  


\subsection{Inverse sequences for circles}
\label{Euclidean-circle}

There are three kinds of geometries that correspond to the intuitive notion of circle understood as closed non-self-intersecting curve with constant curvature. 

The first one is the well known flat circle where the corresponding pattern is  $\mathbb{O}$. 
The second kind consists of elliptic circles that are embedded in Euclidean space. The third kind is for hyperbolic circles that are embodied in hyperbolic plane. 
Can epileptic circles and hyperbolic circles be constructed as intrinsic geometries?

\subsubsection{Hyperbolic circles }

Constructions of hyperbolic circles may use the method used in Subsection \ref{finite-hyperbolic}. The base is the flat circle and the pattern $\mathbb{O}$. Let the points with a fixed finite hyperbolic curvature (like $h_2$) be called hyperbolic points. They can be introduced uniformly in the consecutive graphs in the following way. 
\begin{enumerate}
\item
Starting with the second  graph , 
$[\bar{a}]$ and $[\bar{b}]$ (corresponding to one point) are hyperbolic, as well as $[b.\bar{a}]$ and $[a.\bar{b}]$ (also corresponding to one point) are hyperbolic. Here $[b.\bar{a}]$ means concatenation of $b$ and $\bar{a}$, i.e. $b.a.a.a. \dots$ . Analogously for $[a.\bar{b}]$.  
\item
Starting with the sixth graph,  
$[b.a.\bar{b}]$ and $[b.b.\bar{a}]$ (corresponding to one point) are hyperbolic, as well as $[a.b.\bar{a}]$ and $[a.a.\bar{b}]$ (also corresponding to one point) are hyperbolic. 
\item  
Assuming that $a<_lb$, let us introduce the lexicographical order (denoted $<_l$) on the set $C^{\mathbb{O}}_k$. 
For graph $G^{\mathbb{O}}_k$, and its vertices $x$ and $y$ that are adjacent: 
\begin{itemize}
\item
 If $x <_l y$, then starting with the $2k$-th graph, 
$[x.\bar{b}]$ and $[y.\bar{a}]$ (corresponding to one point) are hyperbolic.  
\item
Otherwise, i.e. if $y <_l x$, then, starting with the $2k$-th graph, $[x.\bar{a}]$ and $[y.\bar{b}]$ are hyperbolic. 
\end{itemize}
\item
And so on. It resembles the construction of an Euclidean circle from regular $2n$-polygons inscribed in the circle. The limit of the polygons, as $n$ goes to infinity, is the circle. 
\end{enumerate}

The hyperbolic points may have different repulsion parameters. So that, the resulting spaces may have hyperboloidal shape.

\subsubsection{Elliptic circles }

Constructions of elliptic circles may be based on the method used in Subsection \ref{finite-elliptic} where one-point elliptic curvature was introduced into flat circle resulting in the space $e\mathbb{O}$. In the very similar way as for hyperbolic circles, elliptic one-point curvatures can be introduced in an uniform way.  The points with such fixed finite elliptic curvature form a dense set in the circle. 


The elliptic points may have different attraction parameters. So that, the resulting spaces may have ellipsoidal shape. 

Also a hybrid construction that uses both hyperbolic points and elliptic points may be interesting.

\subsection{Summary of the examples }

All the above three examples of flat circle, elliptic circle and hyperbolic circle are constructed as inverse sequences of the form  $(G^\mathbb{P}_n, d_n, \phi_{(k+1,k)}; \ n \in N)$. Except the example of infinite hyperbolic one-point curvature from subsection \ref{infinite-hyperbolic}, in the limit as topological spaces, all circles are homeomorphic. From metric point of view, they are isometric. However, from geometrical point of view, they are different. 


The simple examples of elliptic and hyperbolic spaces presented in this section seem to be instructive, and can be easily generalized in order to construct intrinsic sophisticated hyperbolic and elliptic spaces on the base of flat $n$-torus, flat $n$-Klein bottle, and Euclidean $n$-cube. 

Any Euclidean $2$-sphere is elliptic. It is possible (however not so easy) to construct an inverse sequence that results in intrinsic elliptic $2$-sphere. 

Is it possible to construct inverse sequences (intrinsic geometries) that correspond to a flat $2$-dimensional sphere, and hyperbolic $2$-dimensional spheres? 

Flat sphere and hyperbolic sphere contradict our intuition that is evidently based on the Euclidean $3$-dimensional space. The same is true for flat torus and flat circle. 

Do the constructions and the corresponding notions introduced above have any counterparts in our (human) geometrical intuitions? The answer is NO.   The innate intuitions are based on the Euclidean $3$-dimensional space.

A source of non-Euclidean intuitions may be find in general relativity where  gravity is identified with curvature of space.

\section{Curvature}
\label{main-curvature}

Note that the constructions of an intrinsic hyperbolic circle and an intrinsic elliptic circle (presented in the previous section) are based on the intrinsic geometry of the flat circle. 

The idea of curvature arises (in an intuitive sense) if one-point hyperbolic circle as well as one-point elliptic cycle are compared to the flat circle. Since the curvature is a local property, it is  sufficient to compare a geodesic going through a point and belonging to a neighborhood of the point, to a segment of the unit interval of the same length as the length of the geodesic. 

This gives rise to define the notion of parametrization of a geodesic (a curved interval) in order to compute the curvature. 

The notion of curvature (we are going to introduce) is completely different from the geodesic curvature in Riemannian geometry, where it measures how far a curve is from being a geodesic.

\subsection{Parametrization of a geodesic } 

Let $(\mathbb{P}, d)$ be an asymptotic geodesic metric space, and  $(G^{\mathbb{P}},  \phi^{\mathbb{P}}_{(k+1,k)}, d_k; k \in N)$ be the corresponding inverse sequence with metrics. 

Let the space satisfies the property that geodesics are unique locally. This seems to be a crucial assumption to define curvature at a point lying in a geodesic. 

\begin{figure}[h]
	\centering
	\includegraphics[width=0.99\textwidth]{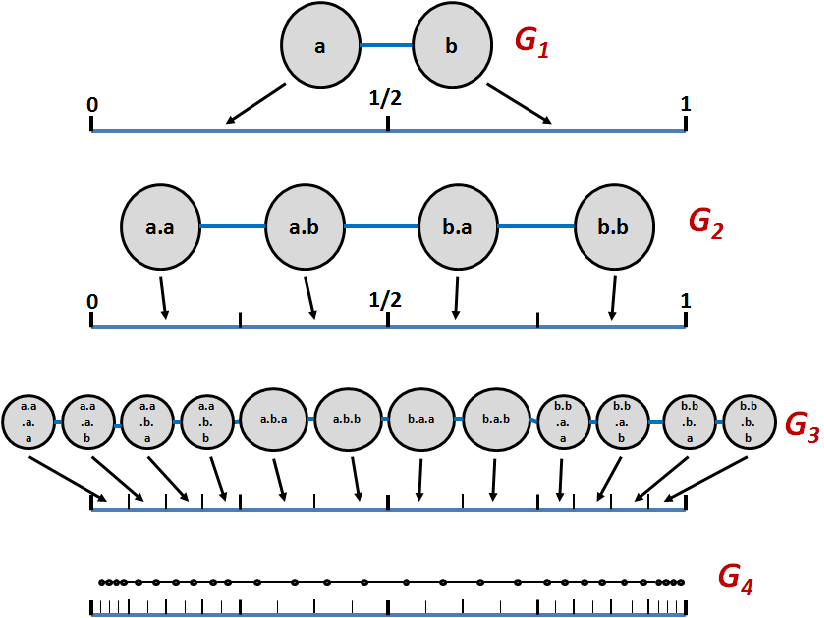}
	\caption{ Pre-parametrization of the geodesic of length $1$ from  $[\bar{a}]$ to  $[\bar{b}]$. The sequence $(G_1, G_2, G_3, G_4, \dots)$ corresponds to the one point hyperbolic circle (see Section \ref{finite-hyperbolic}) where for any $k$ the vertices  $\bar{a}(k)$ and  $\bar{b}(k)$ are not adjacent. Note that in the graphs $G_3$, $G_4$ and $G_k$ in general, metric in a graph is computed as the length of path divided by the graph diameter  }
	\label{curve-parameter}
\end{figure}

Let $\gamma$ be a $(\mathbb{P}, d)$-geodesic, and $(g_k; \ \ k\geq n_\gamma)$ be the sequence of $(\mathbb{P},d_k)$-geodesics that determines it. Equivalently, the geodesic can be considered as the inverse sequence 
($G^{\gamma}_k, \phi^{\gamma}_{(k+1,k)}, d_k; k\geq n_\gamma$), where metric $d_k$ is restricted to the geodesic; see the end of Section \ref{geodesic}. 

Although, any geodesic is homeomorphic to the unit interval, its intrinsic geometry may be locally hyperbolic, or elliptic or flat, see the examples in Section \ref{hyperbolic and elliptic circles}.  

The inverse sequence and the pseudo-metric determine intrinsic geometry of the geodesic. The extrinsic geometry of the geodesic in the ambient space $(\mathbb{P}, d)$ is the same as the intrinsic one. This gives rise to define a new notion of curvature different than the sectional curvature for Riemannian manifolds.  

The new curvature is based on the measurement how a geodesic differs from an Euclidean straight line segment of the same length. 

A sequence $(g_k; \ \ k\geq n_\gamma)$  of $(\mathbb{P},d_k)$-geodesics corresponding to $(\mathbb{P}, d)$-geodesic $\gamma$ determines uniquely 
a bijective mapping  \\ $ \hat{\pi}_{\gamma} : [0; l_{\gamma}] \rightarrow \gamma$, where $l_{\gamma}$ is the length of the geodesic, and $[0; l_{\gamma}]$ is an interval, a subset of $\mathbb{R}$.  

The mapping $ \hat{\pi}_{\gamma}$ (to be defined below) is called a {\em  pre-parametrization of $(\mathbb{P}, d)$-geodesic $\gamma$} for a sequence $(g_k; \ \ k\geq n_\gamma)$ of $(\mathbb{P},d_k)$-geodesics.  

We may imagine that the pre-parametrization describes a point particle moving along the geodesic from its beginning at time $t=0$ to the end of the geodesic at time $t=l_{\gamma}$.

The first derivative of the mapping $\hat{\pi}_{\gamma}$ at time $t$ may be interpreted as an approximation of the velocity of the point  particle at $t$.  The very velocity may be interpreted as a curvature at point $ \hat{\pi}_{\gamma}(t)$ along the geodesic. 

The idea is explained in Fig. \ref{curve-parameter}. Here, a sequence of partitions of the unit interval is constructed from the one-point hyperbolic circle (see Fig. \ref{circle-flat-eliptic-hiperbolic}) by making the vertices $\bar{a}(k)$ and $\bar{b}(k)$ not adjacent for any $k\in N$. The graph $G_k$ determines partition $\Pi_k$ of the unit interval, such that each vertex $x$ of $G_k$ has its own element of the partition. In the next partition $\Pi_{k+1}$, the element is divided into sub-intervals of the same size (any edge has the same normalized length).  
The number of the sub-intervals is equal to the number of successors of $x$ that belong to $G_{k+1}$.  

The key assumption (see Section \ref{Asymptotic pseudo-metric}) is that  the length of each edge (distance between two neighboring vertices in graph $G_k$) is one and the same for all edges and is the minimal distance in $G_k$ according to the metric $d_k$. 

Below we present the construction of the mapping $ \hat{\pi}_{\gamma}$.  

Let us recall that $\gamma$ is identified with the sequence $(g_k; \ \ k\geq n_\gamma)$ of $(\mathbb{P},d_k)$-geodesics.  

For any $x$ in $g_k$ (actually in $G^{\gamma}_k$),  let  $ext^{\gamma}(x)$ denote the maximal sub-path $p=(y_1, y_2, \dots , y_{m_x})$ of $g_{k+1}$ such that for any $y$ in $p$: \  $y(k) = x$, i.e. $\phi^{\gamma}_{(k+1,k)}(y) = x$.

For the sake of presentation, let $n_\gamma = 1$. 

Let us construct the sequence $\Pi^{\gamma} = (\Pi_1, \Pi_{2}, \Pi_{3}, \dots )$ of nested partitions of the interval $[0,l_{\gamma}]$. 
\begin{enumerate}
\item
For $k=1$. Let $m$ denote the number of elements of the sequence $g_k$.  So that, $g_k = (x_1, x_2, \dots, x_{m})$. If $m=1$, then  $\Pi_1$ consists of one element  $[0,l_{\gamma}]$. 

If $m\geq 2$, then the partition $\Pi_1$ of $[0,l_{\gamma}]$ is defined as the set $\{ I_{x_1}; I_{x_2}; \dots ; I_{x_{m}}\}$ of the sub--intervals  of the interval $[0,l_{\gamma}]$, where  

$$
I_{x_1}: = [0,\ \  \frac{l_{\gamma}}{m}]
$$

$$
I_{x_2}: = [\frac{l_{\gamma}}{m},\ \  2\frac{l_{\gamma}}{m}]
$$

$$
I_{x_3}: = [2\frac{l_{\gamma}}{m}, \ \ 3\frac{l_{\gamma}}{m}]
$$

$$
\dots
$$

$$
I_{x_{m-1}}: = [(m-2)\frac{l_{\gamma}}{m}, \ \ (m-1)\frac{l_{\gamma}}{m}]
$$

$$
I_{x_{m}}: = [(m-1)\frac{l_{\gamma}}{m} ,\ \ l_{\gamma}]
$$ 

The mapping from elements of $g_k$ (actually, from $C^{\gamma}_k$ the set of vertices of graph $G^{\gamma}_k$) onto $\Pi_1$ is defined by the correspondence  $x_i \mapsto I_{x_i}$, for $i = 1,2, \dots , m$. Let the mapping be denoted by $f_1$. 
\item 
Suppose that for a number $k \geq 1$ the partition $\Pi_{k}$ and the mapping  
$f_k: g_{k} \rightarrow \Pi_{k}$ have been already constructed. 
 
For any $y$ in $g_k$, $f_k(y)$ is a sub-interval (of $[0, \ \l_{\gamma} ]$)  denoted by $[{\bf 0}_y,{\bf 1}_y]$ and its length is denoted $l_y$.  

Let $n$ denote the number of elements of the sequence $ext^{\gamma}(y)$. So that, $ext^{\gamma}(y) = (y_1, y_2, \dots, y_{n})$. 

Partition $\Pi^{y}_{k+1}$ is defined as the set $\{ I_{y_1}; I_{y_2};  \dots; I_{y_{n}} \}$ of sub--intervals  of the interval $[{\bf 0}_y,{\bf 1}_y]$, where 

$$
I_{y_1}: = [{\bf0}_y,\  \frac{l_y}{n}]
$$

$$
I_{y_2}: = [\frac{l_y}{n},\ \ 2\frac{l_y}{n}]
$$

$$
I_{y_3}: = [2\frac{l_y}{n},\ \ 3\frac{l_y}{n}]
$$

$$
\dots
$$

$$
I_{y_{n-1}}: = [(n-2)\frac{l_y}{n},\ \ (n-1)\frac{l_y}{n}]
$$

$$
I_{y_{n}}: = [(n-1)\frac{l_y}{n},\ \ {\bf 1}_y]
$$ 
 
Let us define the mapping $f^y_{k+1}: ext^{\gamma}(y) \rightarrow  \Pi^{y}_{k+1}$ such that for any $y_i \in ext^{\gamma}(y)$, where $i=1,2, \dots  n$, 
 $$
 f^y_{k+1}(y_i) :=  I_{y_i}
 $$ 
 
Let  $\Pi_{k+1} := \bigcup_{y\in g_k} \Pi^{y}_{k+1}$.   

The mapping   
$f_{k+1}: g_{k+1} \rightarrow \Pi_{k+1}$ is defined as the union of all   $f^y_{k+1}$ for  $y$ in $g_k$. It is a bijective mapping. 
\end{enumerate}

For any $k$, each element of  $g_{k}$ corresponds (via $f_{k}$) to an element of $\Pi_{k}$ and vice versa. Let this correspondence be denoted by $f_{\gamma}$. 

By the above construction, for any thread $u$ of ($G^{\gamma}_k, \phi^{\gamma}_{(k+1,k)}; k\geq n_\gamma$), and for any $k$: 
$$
f_{\gamma}(u(k + 1)) \subseteq  f_{\gamma}(u(k))
$$
 So that, any thread $u$ corresponds to a sequence of nesting sub-intervals converging to a point $r$ in the interval $[0,l_{\gamma}]$. Let this correspondence $u \mapsto r$ be denoted by $f_{\gamma}^{\infty}: C^{\gamma} \rightarrow [0,l_{\gamma}$], where $C^{\gamma}$ is the set off all threads of ($G^{\gamma}_k, \phi^{\gamma}_{(k+1,k)}; k\geq n_\gamma$). 

It is clear that if $[u] = [v]$ (relatively to the equivalence relation $\sim_{d}$), then $f_{\gamma}^{\infty}(u) = f_{\gamma}^{\infty}(v)$. 

Finally, we can define the desired pre-parameterzation $ \hat{\pi}_{\gamma} : [0,l_{\gamma}] \rightarrow \gamma$ as follows
$$
\hat{\pi}_{\gamma}^{-1}([u]) := f_{\gamma}^{\infty}(u)
$$

Note that the definition of the above pre-parametrization of a geodesic $\gamma$ depends on the number $n_\gamma$, i.e. on the initial graphs of the inverse sequence ($G^{\gamma}_k, \phi^{\gamma}_{(k+1,k)}; k\geq n_\gamma$). Since geodesics are limits of their approximations in the consecutive graphs, they do not depend on the initial graphs. This is the very reason to call $ \hat{\pi}_{\gamma}$ a pre-parameterization. The limit, as $n_\gamma$ goes to infinity, is the  {\em natural parametrization} of $\gamma$.  This is not enough for our purpose. The proper parametrization that captures the curvature of a geodesic will be defined in the subsection \ref{parametrization}. 


\subsection{Derivative of pre-parametrization}

Let $\gamma$ be a geodesic in an asymptotic metric space $(\mathbb{P}, d)$, and  
$ \hat{\pi}_\gamma: [0,l_{\gamma}] \rightarrow \gamma $ be the pre-parametrization of $\gamma$. 

Mapping $ \hat{\pi}_\gamma$ can be differentiated. The first derivative may be interpreted as an approximation of the velocity of a point particle traveling along the geodesic within the time interval $[0, l_{\gamma}]$. 

Let $t$ belong to the interior of the interval $[0, l_{\gamma}]$, i.e. $t\in (0, l_{\gamma})$. 
The derivative of 
 $ \hat{\pi}_{\gamma}$ at $t$  denoted  $ \hat{\pi}_{\gamma}'(t)$, is defined as follows. 

 $$
 \hat{\pi}_{\gamma}'(t) := 
 \lim_{s \to 0} \frac{d( \hat{\pi}_{\gamma}(t + s), \hat{\pi}_{\gamma} (t))}{| s |}
 $$
if the limit exists. 

Note that since $ \hat{\pi}_{\gamma}$ depends on the initial graphs in the inverse sequence, also the same concerns $
 \hat{\pi}_{\gamma}'$. 
 
In order to remove the dependence we proceed as follows.

\subsection{Curvature at a point along a geodesic }
\label{curvature}

Let $\gamma_r $ be a geodesic such that $r$ is the middle point of $\gamma_r$. Recall that $l_{\gamma_r}$ denotes the length of the geodesic. 

Let    
$(\gamma^n_r; n\in N)$ be a sequence of geodesics satisfying the following conditions.  
\begin{itemize}
\item  
 $\gamma^{n+1}_r\subset \gamma^n_r\subset \gamma_r$, 
\item
 $l_{\gamma^n_r} < \frac{1}{n}$, 
\item 
$r$ is the middle point of $\gamma^n_r$. 
\end{itemize} 

The curvature at point $r$ in the direction $\gamma_r $ (denoted by $\daleth(r,\gamma_r)$ ) 
is defined as follows.

For any $n\in N$, let $t_n$ be such that $ \hat{\pi}_{\gamma^n_r}(t_n)=r$. 
 
Then, 
$$
\daleth(r,\gamma_r) := 
\lim_{n\to \infty} 
 \hat{\pi}_{\gamma^n_r}'(t_n)
$$
if the limit exists. 


Note that by the definition, $\daleth(r,\gamma_r) = \daleth(r,\gamma^n_r)$ for any $n$. Hence, the curvature does not depend on the initial graphs of the inverse sequence. If the limit exists, then the curvature is also independent on the choice of geodesics $\gamma^n_r$; the only requirement is that their limit is $r$, the common middle point of the geodesics.  
Hence, if $r$ is in the interior of $\gamma$, then  $\daleth(r,\gamma)$ is well defined. Note that $\daleth$ is defined for a fixed asymptotic geodesic  metric space $(\mathbb{P}, d)$ so that, it should be indexed by $(\mathbb{P}, d)$, that is, $\daleth^{(\mathbb{P}, d)}$.   

Flat geometrical spaces like flat circle, flat torus and flat Klein bottle have curvature $1$ at any point. 

It is easy to calculate that the curvature for the space $e\mathbb{O}$ (see Section \ref{finite-elliptic}) at the point $[\bar{a}]=[\bar{b}]$ is $\frac{1}{2}$, whereas  for the space $h_2\mathbb{O}$ (see Section \ref{one-point-hyperbolic}), the curvature is $2$. 

Note that the sectional curvature is not defined for one-dimensional  Riemannian manifolds. 

An asymptotic geodesic metric space $(\mathbb{P}, d)$ is called {\em 1-class smooth}, if the curvature is well defined for all geodesics $\gamma$ and $r\in \gamma$, and is a continuous mapping.

\subsection{Parametrization of a geodesic }
\label{parametrization}

Let an asymptotic geodesic metric space $(\mathbb{P}, d)$, be 1-class smooth. 

For a geodesic $\gamma$ and its pre-parametrization $\hat{\pi}_{\gamma}: [0,l_{\gamma}] \rightarrow \gamma $, let us define the beginning point $r_0 := \hat{\pi}_{\gamma}(0)$ and the end point $r_\gamma :=  \hat{\pi}_{\gamma}(l_{\gamma})$ of geodesic $\gamma$. 

For any $r\in \gamma$, let 
$$
 d_{\gamma}(r) := d(r_0, r)
$$
be the length from the beginning of the geodesic (i.e. from $r_0$) to the point $r$ on the geodesic.

Let us define the partition of initial segment of geodesic $\gamma$, i.e. from $r_0$ to $r$, onto  $n$ sub-segments, of the same length 
$\frac{d_{\gamma}(r)}{n}$, represented by the sequence 
$(r_0, r_1, r_2, \dots , r_{n-1}, r_n)$ where $r_{n}=r$.  

That is, for 
 $k = 0, 1, 2 , \dots , n-1$ 
$$
d(r_k, r_{k+1}) = \frac{d_{\gamma}(r)}{n}
$$ 
or equivalently 
$$
r_{k} := d^{-1}_{\gamma}( d_{\gamma}(r_{k-1}) + \frac{d_{\gamma}(r)}{n})
$$

Let $\hat{r}_{k}\in \gamma$ be the middle point between the points $r_{k-1}$ and  $r_k$. Formally, 

$$
\hat{r}_{k} := d^{-1}_{\gamma}( \frac{1}{2}(d_{\gamma}(r_{k-1}) + d_{\gamma}(r_{k}) ))
$$

Let for any $r\in \gamma$
$$
\tau(r) := \lim_{n\to \infty} \frac{d_{\gamma}(r)}{n} \sum^n_{i=1}  \daleth(\hat{r}_i,\gamma)
$$

The {\em parametrization} of $\gamma$ denoted by 
$\pi_{\gamma} : [0, \tau(r_\gamma)] \rightarrow  \gamma  $, where   $r_\gamma$ is the end of the geodesic, 
is defined as follows. 

$$
 \pi^{-1}_{\gamma} (r) :=  \tau(r)
$$
for $r\in \gamma$. 

By the definition, $\pi_{\gamma} (\tau(r_\gamma)) = r_\gamma$. 

For flat circle, i.e. the space $\mathbb{O}$, the first derivative at any point is equal 1. So that, for any geodesic $\gamma$,  $\tau(r_\gamma)$ is equal to the length of the geodesic, i.e. to $l_\gamma$. 

For a geodesic $\gamma$ with a finite constant curvature, say $\kappa$, \  \  $\tau(r_\gamma) = \kappa l_\gamma$. Perhaps it may be interpreted as follows. The elliptic curvature shortens the distance whereas the hyperbolic curvature enlarges the distance. 


%
%
%

\subsection{Discussion and relation to Riemanian manifolds}

Advanced and sophisticated mathematical notions are used in the definition of Riemannian $n$-manifold. 
Roughly, it is a topological space that is locally (at any  point)  homeomorphic to an open subset of $n$-dimensional Euclidean space together with smooth transformations, from point to point, of these open sets. It is  equipped with a positive-definite metric tensor. 
The Riemann curvature tensor measures the extent to which the metric tensor is not locally isometric to that of Euclidean space.
The abstract notions of: topology, homeomorphism, smooth transformations, metric tensor and curvature tensor are not computational in general.  

Our approach is elementary and is based on simple primitive notions of  finite graphs and inverse sequences. Since all of the constructions of geometries are inductive and operate on finite structures, they are fully computational. Let us emphasize once again that geometrical space, as asymptotic geodesic metric space $(\mathbb{P}, d)$, is an abstraction of the inverse sequence $(G^\mathbb{P}_n,  \phi^{\mathbb{P}}_{(n+1,n)}, d_n; \ n \in N)$. All the notions of geodesics and curvature are defined by means of finite structures of the sequence.

That is, the notions of $d_k$ metric and $(G_k, d_k)$-geodesic are defined for any of the graphs in the inverse sequence.  They uniformly (in the limit) approximate the abstract notions  of metric, geodesic, and curvature in abstract space  $(\mathbb{P}, d)$.  

It seems that any compact Riemannian manifold can be constructed as an inverse sequence of finite graphs along with appropriate metrics. 

However, not all  geometrical spaces can be represented as Riemannian manifolds. The local homeomorphism with an Euclidean space is too restrictive. There are interesting and important geometries that are not locally homeomorphic to Euclidean spaces like Lipscomb spaces \cite {Lipscomb}.

\section{Conclusions}

The computational value of the proposed approach to combinatorial constructions of intrinsic geometries seems to be proved. Still a lot of important questions can be asked like the following ones. 

What are finite structures of graphs in an inverse sequence that determine (along with an asymptotic pseudo metric) the Lebesgue covering dimension, homotopy type of the corresponding space, etc.? 


What about the flat and hyperbolic intrinsic geometries of $2$-sphere?

Flat intrinsic geometries of $n$-torus and $n$-Klein bottle  (for $n>1$) can be used to construct sophisticated hyperbolic geometries. How are the constructions related to the classic notions of geometry? 

Is the proposed framework flexible enough to completely capture our intuitions of geometry? \\ 

The following motto of the famous book \cite{Hilbert}  {\em Foundations of Geometry} by David Hilbert may be an inspiration to further investigations.
 
~\\ 
{\em "All human knowledge begins with intuitions,\\
thence passes to concepts and ends\\
with ideas."}\\
Kant, Kritik der reinen Vernunft, Elementariehre,\\
Part 2, Sec. 2.\\

The vibrant debate on the Foundations of Mathematics and the primeval intuitions of the Geometry in the XIX century were summarized by Ernst Mach (1903)  \cite{Mach}. 
The following citation may give a flavor of the debate. 

{\em "Gauss [in Brief von Gauss an Bessel, 9. April 1830 ] expressed the conviction that it was impossible to establish the foundations of geometry entirely a priori,  and who further asserted that "we must in humility confess that if number is exclusively a product of the mind, space possesses in addition a reality outside of our mind, of which reality we cannot fully dictate a priori the laws."} 

Perhaps, it is the right time to come back to the intuitions, and to continue the debate.

\subsection{Final remarks }
According to Immanuel Kant, the notions of space and time constitute the cognitive framework we (humans) use to structure our experience. Spatial measurements are used to quantify distance between objects and their size. Temporal measurements are used to quantify intervals between (or duration of) events. The notions are embedded and innate in our minds implemented as neural networks in human brains. The networks, by their nature, are finite structures. 

Space-time continuum is a mathematical model of the so called {\em real universe}. It is a four-dimensional manifold which combines the three dimensional Euclidean space and the one dimensional real line of time into a single four-dimensional manifold. 

In $n$-dimensional Euclidean space ($\mathbb{R}^n$) any point can be specified by $n$ real numbers. These are called the coordinates of the point.

Any $n$-dimensional Riemannian manifold is a generalization of $n$-dimensional Euclidean space where the coordinates are defined only locally. 

The Continuum (as $\mathbb{R}$ - the set of the real numbers) is an actual infinity. It consists of uncountable number of abstract points. 

A human can perceive and process only finite structures. How, then, the notion of space-time continuum is related to our (human) mental abilities? 

The prevailing and ubiquitous view (paradigm) of the contemporary Mathematics (since the beginning of XX century) is that axiomatic approach is the only right way to investigate sophisticated abstract objects (being actually  infinite) and relations between them, as well as ``create'' new ones whatever they might mean. Theorems and their proofs (supposed to be  expressed in a formal language) are considered as the essence (the results) of Mathematics. 
The meaning of this formal language is ignored. Sometimes it is referenced to the enigmatic {\em intuition} where the abstract objects, their properties, and relations between them (described by the language in the theorems) are supposed to exist.      

In opposition to the current paradigm, we propose a radically different approach to Mathematics. 
Humans are able only to construct inductive structures that are the {\em intuition} - the true concrete essence of Mathematics. 
Usually, an inductive structure is a sequence of finite structures where any consecutive structure (element) of the sequence is constructed on the basis of the previous already constructed elements. Human ability to abstraction is the crucial point in developing abstract objects, their properties and relations between them. 

The basic abstraction is an aggregation of inductive structures that share some properties. Abstraction can be applied hierarchically to yet already defined abstract objects.  The final stage of such abstractions are axiomatic formal theories like group theory, set theory, topology, and category theory. 

However, there is the strict border of such abstractions. An abstract object that itself is actually infinite cannot be treated as a finite structure. 
 
Properties of abstract objects and relations between them  should be always grounded in the properties and relations of inductive structures from which they are abstracted from. 

Natural numbers constitute the basic inductive structure. The set of real numbers is an actual infinity, that became the basic  notion in the contemporary Mathematics since the end of XIX century. Actually, the real numbers are an abstraction of relations on natural numbers. 

The principal goal of the paper was to show that geometries can be constructed as simple abstractions of inductive combinatorial structures. That is, an inductive sequence of finite graphs with some regular properties and metrics, may be abstracted to a geometry. This concerns the classic Euclidean geometry ($\mathbb{R}^n$ spaces), Riemannian manifolds as well as other new geometries. 

The paper is a continuation of our previous work on the ultimate Foundations of Mathematics in the spirit of the Intuitionism of Brouwer. The paper  \cite{TO} is on higher-order mathematical logic (functionals), and \cite{C} is on the Continuum.


\end{document}